\documentclass[12pt]{article}
\usepackage{latexsym}
\usepackage{amsmath}
\usepackage{amsfonts}
\usepackage{mathrsfs}
\usepackage{dsfont}
\usepackage{multibox}
\usepackage{verbatim}
\usepackage{hyperref}
\usepackage{tikz}
\usetikzlibrary{matrix,arrows}
\usepackage{subfigure}

 \csname
@addtoreset\endcsname{equation}{section}
\usepackage{amsmath,amssymb}
\usepackage{graphicx}
\usepackage{color}

\begin{document}

\def\be{\begin{equation}}
\def\ee{\end{equation}}
\def\bea{\begin{eqnarray}}
\def\eea{\end{eqnarray}}
\def\nn{\nonumber}
\def\n{n+\delta'}
\def\barn{\bar{n} + \bar{\delta}'}
\def\m{m+\delta''}
\def\barm{\bar{m}+\bar{\delta}''}

\begin{flushright}

\end{flushright}

\vspace{40pt}

\begin{center}

{\Large\sc Closed String Partition Functions in Toroidal Compactifications of Doubled Geometries}

\vspace{50pt}

{\sc Hai Siong Tan}

\vspace{15pt}
{\sl\small
Institute for Theoretical Physics,
University of Amsterdam, \\
Science Park 904, Postbus 94485, 1090 GL, Amsterdam, The Netherlands
\\

\vspace{10pt}
{\it H.S.Tan@uva.nl}}

\vspace{70pt} {\sc\large Abstract}\end{center}

We revisit partition functions of closed strings on toroidal backgrounds, including their $\mathbb{Z}_N$ shift orbifolds in the formalism where the dimension of the target space is doubled to make T-duality manifest. In such a T-duality covariant formalism, the constraint equation imposes a form of chiral factorization. Our computation 
furnishes a non-trivial consistency check for the quantum worldsheet theory of the doubled sigma model, when strings are placed on general toroidal backgrounds. The topological term that mixes the physical space and its T-dual is crucial in demonstrating that chiral factorization works, and that we obtain the correct partition function after imposing the constraints. Finally, we discuss how our results extend to  $\mathcal{N}=1$ worldsheet supersymmetry and string worldsheets of higher genus. 
\newpage

\tableofcontents
\section{Introduction}\label{sec:Intro}
\label{Intro}

This paper is devoted towards understanding some aspects of the doubled formulation of string theory of C.Hull \cite{Hull:2006va} - a T-duality covariant framework that lives on an enlarged target space in which one can think of the extra coordinates as those conjugate to string winding numbers. Over the years, this formalism has attracted much attention\footnote{
For a sampling of recent literature over the past decade, see for example \cite{DeAngelis,Hohm,Hull:2007,Hull:2009,ReidEdwards,Hull:2004,Dabholkar,Lee}.}
primarily as a possible way to formulate our understanding of a class of non-geometric string backgrounds known as T-folds, itself being based on a series of earlier seminal works in the 80's and 90's (\cite{Cremmer,Tseytlin:1990,Tseytlin:1991,Maharana,Hull:1988}).
String theory can be consistently defined on many non-geometric backgrounds which can nonetheless give rise to a conventional theory endowed with geometric four-dimensional spacetime (see for example \cite{Flournoy}). An interesting class of non-geometric backgrounds are those which are constructed by duality twists \cite{Dabholkar:2002}. In these twisted backgrounds, fibres are glued together with T-duality-valued transition functions - namely, the `T-folds'. More generally, we can have `U-folds' with U-duality-valued transition functions \cite{Kumar}.\footnote{See also \cite{Ganor} for a recent attempt at understanding an S-duality twisted gauge theory, \cite{Ganor:1998} for a study of U-duality twists in toroidal compactifications, and \cite{deBoer} for non-geometric branes that can arise naturally in M-theory. Most recently in \cite{Andriot}, nongeometric fluxes related to duality twists are discussed in the context of supergravity.} 

Recall that T-duality was originally demonstrated to be a symmetry of the string sigma model with toroidal target spaces in which there is a $U(1)^d$ isometry\footnote{See for example \cite{Giveon} for an excellent review.} and thus are principal $U(1)^d$ bundles. One can generalize the notion of T-duality to more generic torus bundles with less symmetry under some conditions, even in spaces devoid of globally defined Killing vectors (see \cite{Hull:2006} for details). This sets up the notion of a T-fold which locally resembles a conventional spacetime patch with a torus fibration over some open set in a base space, with T-duality-valued transition functions. The importance of these backgrounds arises from the fact that applying T-duality to geometric backgrounds with Kalb-Ramond $B$ fluxes can sometimes give rise to T-folds. To be more precise, the moduli of the string CFT which are related via an $O(d,d;\mathbb{Z})$ transformation gives essentially the same CFT, and the transition functions allow one to construct a bundle of torus CFT over some base space. The moduli can depend on the coordinates of the base space in general. Integrating over the base space then yields the quantum string theory of the T-fold. 

In this doubled formalism, we introduce an extra set of target space coordinates $\tilde{X}_i$ which are conjugate to the winding numbers. They can be understood as the coordinates which are T-dual to $X$ by the $O(d,d;\mathbb{Z})$ element $M$ which is the analogue of the $R\rightarrow \alpha'/R$ duality of a circle target space. When the closed string is endowed with the target space metric $G$ and coupled to the Kalb-Ramond field $B$, this T-duality element inverts a generalized background matrix $E = G + B$. The doubled sigma model has an expanded target space $\mathbb{M}$ of which coordinates patch together to form a $T^{2d}$ bundle over some base $N$. One then has to further impose a constraint which demands half of the $2d$ fields on $\mathbb{M}$ to be left-movers and the other half as right-movers. The ordinary theory is recovered after choosing half of $\{ X^i, \tilde{X}^i\}$ to be the physical space. Such a choice of polarization $T^d \subset T^{2d}$ is generally done locally in each patch for an arbitrary well-defined T-fold. Being able to implement the constraint consistently is crucial. In \cite{Hull:2006va}, and as we shall review in Section \ref{Prelim}, it was shown that this constraint requires a certain set of conserved currents $J$ to vanish and this could be imposed by gauging half of $J$'s via an auxiliary gauge field say $C$. Gauge-fixing and integrating out $C$ recovers the ordinary sigma model locally. The polarization then corresponds neatly to the choice of which $J$'s to gauge. 

A crucial ingredient of the doubled sigma model is a topological interaction term included in the Lagragian as
\be
\label{topologicalterm}
i\pi dX \wedge d\tilde{X}
\ee
and it was argued in \cite{Hull:2006va} that at the level of the classical action, the above term is necessary to yield complete equivalence to the usual string sigma model formulation. Note that such a term does not change the classical theory, but it introduces some relative signs when we sum up the cohomological sectors in evaluating the partition function. We will discuss more about this term in Section \ref{Prelim}. Essentially, its presence preserves the invariance of the doubled sigma model's action under large gauge transformations of the auxiliary coordinates. 

Shortly after the appearance of \cite{Hull:2006va}, Berman and Copland explained in an elegant paper \cite{Berman} that the constraints imposed to halve the degrees of freedom of the doubled formalism can be interpreted as those that determine chiral scalars. The one-loop partition function of a doubled $S^1$ target space is first computed without imposing any constraints but they keep track of certain left- and right-moving degrees of freedom encapsulated in each set of holomorphic variables $P$ and anti-holomorphic variables $Q$. After performing a series of Poisson resummations, the partition function can be expressed in the form of a chiral decomposition, i.e. a product of a holomorphic function and an anti-holomorphic function. Each chiral sector is then further decomposed into two separate blocks, and imposing the constraint is argued to be effectively equivalent to discarding one block in each of the two chiral sectors. The most significant observation made in \cite{Berman} is that the one-loop closed string partition function can be reproduced from the doubled formalism via the chiral factorization procedure explained above provided the topological interaction term is included. This then yields a consistency test for the inclusion of this term in the doubled sigma model Langrangian. 

Inspired by the results of \cite{Berman}, we will build on their work and generalize their results to conventional non-T-fold backgrounds. We find that via the procedure of chiral factorization, we can reproduce the closed string partition functions of any worldsheet genus from the doubled sigma model for the generic toroidal background, including a non-vanishing $B$ field, and thus demonstrate explicitly the quantum equivalence to the usual formulation for toroidal backgrounds. We hope that this acts as a starting point for investigating whether generic well-defined T-folds can be understood at least partially by some principle of chiral block decomposition. 

Most of our technical focus will be on reassembling the classical instanton sums in the partition function via Poisson resummations. We will also discuss how chiral factorization works for the quantum part of the partition function that captures all the string's oscilllator modes.  We should mention that throughout our work, by partition function, we refer to the sum at each genus with a fixed worldsheet metric of unit determinant. The complete string partition function of course involves an integral over the moduli space of Riemann surfaces and appropriate Faddeev-Popov ghost field modes.

With regards to the point about holomorphic factorization, we find it useful to work with orbifolds as an anchor for future explorations. For instance, for asymmetric orbifolds, some have observed that the partition functions can be constructed by sewing together chiral blocks of associated symmetric orbifolds \cite{Aoki}.  As the simplest example, in \cite{HackettJones}, a T-fold - as an asymmetric reflection orbifold - was studied in the doubled sigma model formulation. Corresponding to the only non-trivial element of $O(1,1;\mathbb{Z})$, an $S^1$ is fibred over a base $S^1$, with an asymmetric reflection acting on the right-movers augmented with a $\mathbb{Z}_2$ translational shift in the base. In that work, the doubled sigma model is analyzed in the operator formalism, and it was shown that to recover a modular invariant partition function, one needs to impose the eigenvalues of the T-duality operator to be of certain values. Such a rule was derived in \cite{Simeon}, but we do not know how it could be derived from first principles in the doubled formulation. 

In our work, we will initiate an analysis along a related direction. We consider $\mathbb{Z}_N$ orbifolds with the discrete orbifold action being independent translations in the left and right movers. Apart from being a simple class of orbifolds in its own right, such an orbifold action arises naturally in T-folds which admits description as asymmetric orbifolds. In these backgrounds which are sometimes called `interpolating orbifolds' \cite{Williams}, the absence of singularities necessitates the orbifold action on the fibres to be augmented by a corresponding shift in the base. As an example where we take the base to be $S^1$, we can write the orbifold action as \cite{Narain}
$$
\mathcal{G} = e^{2\pi (F_L + F_R)}\delta_{a,b}
$$
where $\delta_{a,b}$ is an order-$N$ shift along the base circle and $F_{L,R}$ are generators of the orbifold rotation on the fibre coordinates. If $F_L \neq F_R$, this is an asymmetric orbifold \cite{Narain,Narain:1990}. The shift along the base $S^1$ can be parametrized by a vector $\lambda = \left( \frac{a}{N}, \frac{b}{N} \right)$ with $a$ and $b$ defined modulo $N$. Let $X$ denote the base circle coordinate and $\tilde{X}$ its T-dual. Then $\delta$ acts as 
\be
\delta: X \rightarrow X + a\frac{2\pi R}{N},\qquad \delta: \tilde{X} \rightarrow \tilde{X} + b\frac{2\pi}{NR}
\ee
In Section \ref{Prelim}, we shall review these translational or shift orbifolds $S^1/\mathbb{Z}_N$ with slightly more details. Formally speaking, we first need to specify in principle what defines the doubled orbifold that descends correctly to the physical one. A natural rule (that is assumed also in \cite{HackettJones}) is that $\tilde{X}$'s are acted upon by an operator equivalent to the T-dual (defined by the element $M$) of the orbifolded $X$. Thus, $\mathbb{Z}_N$ translational/shift orbifolds is perhaps the most straightforward to study as a first example, since the action on $X$ and $\tilde{X}$ are just translations along each homological cycle of the worldsheet. Since $\tilde{X} \sim X_L - X_R$, an asymmetric orbifold is obtained when we specify a non-zero shift for $\tilde{X}$. Nonetheless, there is an important caveat that we should mention: namely that not all the asymmetric shift orbifolds on their own enjoy modular invariance which might demand among other things a certain number of spacetime dimensions (see \cite{Aoki} and \cite{Simeon} for a nice discussion on this). We reserve discussion of this important issue to future work \cite{future}, while maintaining our focus on studying the equivalence between the doubled formalism and the conventional sigma model.  

We find that when strings are placed on an orbifold defined by translational shifts in $X$ and $\tilde{X}$, the signs induced by the topological term do lead us to the correct string partition function after chiral factorization. For a non-geometric asymmetric action, where there is an orbifold action on $\tilde{X}$, in the Lagrangian representation of the partition function, this amounts to multiplying $e^{-S_{cl.}}$ with phase factors that parametrize the shifts on $\tilde{X}$. This can be related after a Poisson resummation to the fact that while the momentum generates translations in $X$, the winding number operator generates translations in $\tilde{X}$, and the latter manifests itself as phase factors. The fact that we can establish the equivalence of the partition functions rather easily (as we shall observe in this entire work) points swiftly to the underlying fact that in the doubled sigma model, $X$ and $\tilde{X}$ represent the Fourier transform duals of the momentum $(n)$ and winding numbers $(w)$. Thus, as explained in \cite{Hull:2006}, we can write down the dual state
\be
\label{TdualityFourier}
|X,\tilde{X} \rangle = \sum_{n,w} e^{\frac{inX}{R}} e^{iw\tilde{X}R} |n,w \rangle
\ee
where the dual $\tilde{X}$ corresponds to the T-dual of $X$ in which all original radii are inverted. In the doubled Lagrangian, the natural metric on the fibre is given by $H$ (see eqn. \eqref{Hmatrix}), but this is precisely the matrix which defines a norm for $(n\,\,w)$ that yields the string spectrum, i.e.
\be
\frac{1}{2} M^2 = (n\,\, w) H (n\,\, w)^T
\ee
We will see that our results for the $\mathbb{Z}_N$ shift orbifolds turn out to be very compatible with such a picture. 

The plan of our paper is as follows. In Section \ref{Prelim}, we briefly review the doubled geometry formalism of Hull and some basic aspects of the $\mathbb{Z}_N$ shift orbifolds. In Section \ref{Sorbifold}, we demonstrate the equivalence of the partition function with that obtained in the conventional theory, following closely the technique introduced in \cite{Berman} and \cite{Moore}. At the same time, we present a slightly more efficient way to obtain the partition function which we will employ in the subsequent sections, but if desired, the reader can repeat the method in Section \ref{Sorbifold} for the latter computations and obtain identical conclusions too. Then in Section \ref{Toroidal}, we recover the partition function of the general toroidal background. We find that our narrative can be extended straightforwardly to higher-genus string worldsheets which we study explicitly in Section \ref{Genus}. In Section \ref{Quantum}, we discuss how chiral factorization works for the `quantum part' of the partition function - the part that counts all the oscillators' modes which is effectively the determinant of the Laplacian operator.
We have included a short appendix that touches on some pedagogical background relevant to Section \ref{Quantum} for the sake of readers unfamiliar with Schottky uniformization of Riemann surfaces. Before we end with a concluding discussion, we briefly discuss the case of $\mathcal{N}=1$ worldsheet supersymmetry in Section \ref{SUSY}. 

We emphasize that our various computations pertain not to the bundle structure of a T-fold with non-trivial $O(d,d;\mathbb{Z})$ transition functions, but simply a doubled toroidal geometry. A global choice of polarization can thus be made. We hope that this paper forms a starting point for future explorations of the role of chiral block decomposition in understanding T-folds in the doubled formalism \cite{future}. Throughout our paper, we take $\alpha'=1$.


\section{Preliminaries}
\label{Prelim}

\subsection{On the action of the doubled sigma model}

In the following, we shall give a quick review of the doubled sigma model following \cite{Hull:2006va}. In this T-duality covariant framework, the toroidally compactified dimensions are doubled, and thus so are the number of degrees of freedom. This necessitates introducing constraints to remove the auxiliary ones. In its original formulation that involves a $T^{2d}$ bundle over some toroidal base of which coordinates we shall denote as $Y^n$, the Lagrangian reads
\be
\label{doubledL1}
\mathcal{L} = \frac{\pi}{2} H_{ij} \partial_a \mathbb{X}^i \partial^a \mathbb{X}^j + \pi \partial_b \mathbb{X}^i 
\left( A_{in} \partial^b Y^n + \bar{A}_{im} \epsilon^{bc}  \partial_c Y^m \right) + \mathcal{L}_{base}(Y)
\ee 
where $\mathbb{X}$ are the fibre coordinates,
and the worldsheet one-forms $A^i \equiv A^i_m dY^m$ play the role of a connection. They are the conserved currents $J\equiv A^i_m dY^m + \bar{A}^i_n \ast dY^n$ under $M \in O(d,d;\mathbb{Z})$ transformations which leaves \eqref{doubledL1} invariant under
$$
\mathbb{X} \rightarrow M \mathbb{X},\qquad H \rightarrow (M^{-1})^T H M^{-1},\qquad J \rightarrow (M^{-1})^T J
$$
where the generalized metric reads
\be
\label{Hmatrix}
H \equiv \left( \begin{array}{cc} 
G - BG^{-1}B & BG^{-1} \\
-G^{-1}B & G^{-1}
\end{array} \right).
\ee
The constraint equation reads
\be
\label{constraint}
\ast \partial \mathbb{X}^i = L^{ik} H_{kj} \partial\mathbb{X}^j + L^{im}J_m 
\ee
where in a certain basis, the $O(d,d;\mathbb{Z})$ metric reads
$$
L_{ij} = \left(   \begin{array}{cc} 0 & \mathds{1}_d \\ \mathds{1}_d & 0 \end{array} \right).
$$
One also needs to choose some splitting of $T^{2d}$ into a physical $T^d$ and a dual $\tilde{T}^d$ for each point in the $T^{2d}$ bundle, thus dividing the fibre coordinates into a physical $T^d\, (X)$ and a dual $\tilde{T}^d (\tilde{X})$. The theory can then be written in terms of $X$ alone after solving for $\tilde{X}$ in terms of $X$. 
In this paper, we shall restrict ourselves to backgrounds in which a global polarization can be chosen. We can set the connection to be zero, and the constraint equation then reads
\be
\label{constraintFlatconnection}
\left( \begin{array}{c} \ast \partial X^i \\ \ast \partial \tilde{X}^i \end{array} \right) =
 \left( \begin{array}{cc} 
-G^{-1}B  & G^{-1} \\
G-BG^{-1}B & BG^{-1}
\end{array} \right) 
\left( \begin{array}{c} \partial X^i \\ \partial \tilde{X}^i \end{array} \right) 
\ee
Diagonalizing the matrix $LH$ above, we find that we can express the constraint in terms of self-dual and anti-self-dual equations for the fields (with components in a flat tangent space)
\bea
\label{PQdefinition}
P^a &=& \frac{1}{\sqrt{2}} \left[ \left( e^a_i - B_{ij}e^{ja} \right) dX^i + e^{ja} d\tilde{X}_j \right], \cr
Q^a &=& \frac{1}{\sqrt{2}} \left[ \left( e^a_i + B_{ij}e^{ja} \right) dX^i - e^{ja} d\tilde{X}_j \right], 
\eea
where we have introduced the vielbein by the relation $G^{ij} = e^i_a e^j_b \eta^{ab}$. On the worldsheet, this implies the `chirality' condition\footnote{We note that when the total spacetime dimension is just one, this reduces to the expressions found in \cite{Berman}, namely 
$P = \frac{1}{\sqrt{2}} \left( R dX + R^{-1} d\tilde{X} \right),\,\, Q = \frac{1}{\sqrt{2}} \left( RdX - R^{-1} d\tilde{X} \right). $
}
\be
\partial_{\bar{z}}P^a = \partial_z Q_a = 0
\ee
We find that the doubled Lagrangian can be written in terms of these variables as the sum of two kinetic terms
\be
\label{PQaction}
L_{PQ} = \frac{\pi}{2} \eta_{ab} dP^a \wedge \ast dP^b + \frac{\pi}{2} \eta^{ab} dQ_a \wedge \ast dQ_b +\mathcal{L}_{top.}
\ee
Later in Section \ref{Quantum}, we shall see that for chiral factorization to work, it is rather important that the Lagrangian can be expressed in this manner. Note that there is also a topological term that we have added to \eqref{PQaction} which we shall briefly review next. Recall that since we are restricting ourselves to globally well-defined splitting of the fibre in the doubled torus bundle, we set the $O(d,d;\mathbb{Z})$ connection to be zero. The conserved currents $J_I$ read
\be
\label{current}
J_I = H_{IJ} \partial \mathbb{X}^J - L_{IJ} \ast \partial \mathbb{X}^J \equiv \mathcal{J}_I - j_I
\ee
where $\mathcal{J}_I$ is the Noether current and $j_I$ is trivially conserved. The field equations imply that $d \ast J_I = 0$ but the constraint equation sets the stronger condition $J_I=0$. The current $\mathcal{J}_I$ is the Noether current for the translational symmetries along the fibres which we parametrize as $\delta \mathbb{X} = \alpha$. If we wish to write it with regards to left- and right-moving degrees of freedom, then the symmetries are $\delta X_R = \alpha_R, \delta X_L = \alpha_L$, with $J^i$ generating the anti-diagonal subgroup $\alpha_L = -\alpha_R$, while $J_i$ generates the diagonal subgroup $\alpha_L = \alpha_R$. To impose it, we gauge the symmetry by introducing a gauge field $C$ which is a one-form on the worldsheet.  We perform a minimal coupling procedure which amounts to replacing 
\be
\label{minimalc}
\partial \tilde{X}_i \rightarrow \partial \tilde{X}_i + C_i
\ee
Since this generates the linear coupling term$\sim C_i \wedge \ast H_{ij} \partial X^j$, we add a term 
\be
\label{additionalterm}
C_i \wedge \partial X^i
\ee
to yield the coupling $C\wedge \ast J$. Under the gauge transformations $\delta \tilde{X}_i = \alpha_i, \delta C_i = -\alpha_i$, the action is invariant up to a surface term. To remove this term, we add the topological term to the action
\be
\label{topological}
\mathcal{L}_{top.} = - i\pi d\tilde{X}_i \wedge dX^i
\ee

In \cite{Hull:2006va}, it was noted that since after the inclusion of \eqref{minimalc}, \eqref{additionalterm} and \eqref{topological}, the doubled lagrangian can be written as 
\bea
\label{actionHull}
\mathcal{L} &=& \pi G_{ij} \partial X^i \wedge \ast \partial X^j + \pi B_{ij} \partial X^i \wedge \partial X^j + \frac{\pi}{2} G^{ij} D_i \wedge \ast D_j,\cr
D_i &\equiv& C_i + d\tilde{X}_j - G_{ij} \ast \partial X^j - B_{ij} \partial X^j
\eea
Imposing the constraint is then equivalent to treating $D_i$ as the non-dynamical fields. The invariance under $\delta \tilde{X}_i = \alpha_i$ implies that $\tilde{X}$ can be completely gauged away including winding modes. 

In this paper, we shall begin with the action by gauge-fixing $C$ to be zero, with thus only the topological term being added to the action \eqref{PQaction}. The non-dynamical degrees of freedom are removed from the partition function of the doubled sigma model by identifying and retaining the appropriate holomorphic and anti-holomorphic pieces following the constraint, rather than integrating over the auxiliary fields $D$ in \eqref{actionHull}. This procedure is sufficient for our purpose of reducing the partition function to the correct one, but would certainly need to be modified for T-folds in general. In these cases, the $O(d,d;\mathbb{Z})$ connection should enact an integral role since they measure the non-trivial nature of the bundle.

\subsection{On $S^1/\mathbb{Z}_N$ orbifolds: some generalities}
In this subsection, we shall briefly discuss a somewhat different topic - closed string orbifolds\footnote{See for example \cite{Dixon,Morozov}.}, in anticipation of Section \ref{Sorbifold} where we discuss the $S^1/\mathbb{Z}_N$ case in detail. In particular, non-geometric backgrounds can be obtained when the orbifold action is asymmetric. Let us begin with the simplest example of a $S^1$ target space geometry on a toroidal worldsheet parametrized by periodic variables $\sigma_{1,2} \sim \sigma_{1,2} + 1$ and equipped with the following metric 
$$
ds^2 = \frac{1}{\tau_2} \vert d\sigma_2 + \tau d\sigma_1 \vert^2
$$
where $\sigma_1, \sigma_2$ are the periodic Euclidean time and space worldsheet coordinates respectively. 
Let the periodic boson be $X$, with $R$ being the compactification radius, and its T-dual be $\tilde{X}$. 
In the following, we will absorb $R$ in the target space metric, and let $X$ be a dimensionless coordinate with unit period. First, let us recall some basic facts about abelian orbifolds defined by $\mathbb{Z}_N$ shifts
of the form 
\be
\label{symmetricOrbifold}
\hat{s} \, X \rightarrow X + \frac{1}{N}.
\ee
With regards to the left- and right-movers, the orbifold action acts symmetrically on both, with each of $X_L$ and $X_R$ being shifted by $\frac{1}{2N}$. The partition function can be written as a sum of partition traces
\be
\label{partitionfunctiongeneralform}
Z_{S^1/\hat{s}} = \frac{1}{N} \sum_{g,h=1}^{\hat{s}^{N-1}} {Z^g}_h
\ee
where in the above notation, $h$ labels the twisted sector and
$g$ is defined as the orbifold twist operator (in this case, a translation operator) inserted in the Hilbert space trace. We can describe the shifted boundary conditions by the characteristics $\delta = (\delta', \delta'')$ where $\delta', \delta'' = 0,1/N,2/N,\ldots, (N-1)/N$. Corresponding to $h = \hat{s}^{N\delta'}$ and $g = \hat{s}^{N\delta''}$, the twisted boundary conditions of $X$ read
\bea
\label{periodicityconditions}
X(\sigma_1 + 1, \sigma_2) = X(\sigma_1, \sigma_2) + \delta'' \qquad (\textrm{mod}\,\,1)\cr
X(\sigma_1, \sigma_2 + 1) = X(\sigma_1, \sigma_2) + \delta' \qquad (\textrm{mod}\,\,1).
\eea
The classical zero modes have winding numbers along each worldsheet direction which we denote by $m,n$. Explicitly, we write
\be
X^{cl.}_{m,n} (\sigma) = \sigma^1 (m + \delta'') +  \sigma^2 (n + \delta').
\ee
The string path integral, with Lagrangian $\pi R^2 \partial^m X \partial_m X$ can be split up into a product of a quantum part capturing degrees of freedom of $X-X^{cl.}$ coming from all the oscillators modes, and the classical zero modes (or string `instantons') of which contributions read (we set $\alpha'=1$)
\be
\sum_{m,n} e^{-\frac{\pi R^2}{\tau_2}  \lvert \tau(n + \delta') - (m + \delta'') \rvert^2    }.
\ee
Together with the quantum part, the partition function 
in each twisted sector then reads 
\be
{Z^g}_h (\tau) = \frac{R}{\sqrt{\tau_2}|\eta(\tau)|^2} \sum_{m,n} e^{-\frac{\pi R^2}{\tau_2} \lvert \tau(n + \delta') - (m + \delta'') \rvert^2 }
\ee
where the factor $R$ arises from integrating over the volume of the target space. There is another way to express the above partition function in terms of the chiral block structure of the theory, namely the left- and right-moving string degrees of freedom. One simply needs to perform a Poisson resummation in $n \rightarrow w$, using the formula\footnote{This is a special case of 
\bea
\label{Poissongen}
&&\sum_{m^1,\ldots, m^n \in \mathbb{Z}} \textrm{exp} \left\{ - \pi A_{ij}(m^i + C^i)(m^j + C^j) 
+2\pi i B_i (m^i + C^i) \right\} \cr
&&\qquad = \frac{1}{\sqrt{\textrm{det}\,A}} \sum_{ m_1,\ldots, m_n \in \mathbb{Z}}
\textrm{exp} \left\{ - \pi (A^{-1})^{ij} (m_i - B_i)(m_j - B_j) + 2\pi i C^i (m_i - B_i) \right\}
\eea
where $A$ is an invertible $n \times n$ symmetric matrix and $B,C$ are two $n$-column vectors. 
}
\be
\label{Poisson}
\sum_n \textrm{exp} \left( - \pi A (n+ c)^2 \right) = \frac{1}{\sqrt{A}} \sum_w \textrm{exp} \left[ - \frac{\pi}{A}w^2 + 2\pi i cw \right]
\ee
which gives us
\be
\label{momentumshift}
{Z^g}_h (\tau) = \textrm{Tr}_h \left( g q^{\frac{1}{4} p^2_L} \bar{q}^{\frac{1}{4} p^2_R} \right)=
\frac{1}{|\eta|^2} \sum_{n,w}  e^{-2\pi i \delta'' w}
q^{\frac{1}{4} \left( \frac{w}{R} + R(n + \delta')  \right)^2} 
\bar{q}^{\frac{1}{4} \left( \frac{w}{R} - R(n + \delta')  \right)^2}.
\ee
In \eqref{momentumshift}, we see that in the basis labelled by momentum number $w$ and winding number 
$n$, $g = \hat{s}^{N\delta''}= e^{-i(p_L+p_R)\delta X}=e^{-2 \pi i \delta'' w}$. More generally, one can set up an orbifold by independent shifts in $X$ and its T-dual $\tilde{X}$ which has an inverted compactification radius. For definiteness, let us consider the case where the shifts are 
\be
X \rightarrow X + a\frac{1}{N},\qquad \tilde{X} \rightarrow \tilde{X} + b\frac{1}{N}
\ee
where $a,b=\left\{0,\pm 1\right\}$ are defining parameters of the translation orbifolds\footnote{See \cite{condeescu} for a recent work in which these orbifolds are investigated from the viewpoint of gauged supergravities.}, so for example, at the self-dual point $R=1$, the orbifold of $(a,b)=(1,1)$ would correspond to $X_R \rightarrow X_R, X_L \rightarrow X_L + \frac{1}{N}$. We should introduce two other shift parameters $\bar{\delta}, \bar{\delta}''$, and thus altogether we have
$
\delta, \delta'' = 0,1/N, \ldots, a(N-1)/N\,\,\,\, \textrm{and}\,\,\,
\bar{\delta}, \bar{\delta}'' = 0,1/N, \ldots, b(N-1)/N.\,\,\,\,
$
In each twisted sector labeled by $\delta', \delta''$, the instanton part of the partition function which corresponds to summing over all classical backgrounds with different winding modes reads 
\be
\label{orbifoldZ}
Z^{(\delta'', \bar{\delta}'')}_{(\delta',\bar{\delta}')} = \sum_{w,n} e^{-2\pi i \delta''(w+\bar{\delta}') -2\pi i \bar{\delta}''(n+\delta')  } q^{\frac{1}{4} \left( \frac{w + \bar{\delta}'}{R} +   R(n + \delta')  \right)^2} 
\bar{q}^{\frac{1}{4} \left( \frac{w + \bar{\delta}'}{R} -   R(n + \delta') \right)^2}.
\ee
Note that \eqref{momentumshift} corresponds to the case of $(a,b)=(1,0)$. After a T-duality, we obtain the case of $(a,b) = (0,1)$ which shifts the T-dual coordinate $\tilde{X}$. This can be called a winding number shift, dual to the previous momentum shift. More generally, we can have both simultaneously. For example, when $a=b=1$ or whenever $b\neq 0$, we have an asymmetric orbifold. This presents a rather natural setting for the doubled formalism where the parent sigma model includes $\tilde{X}$ as a bona fide target space coordinate right from the outset. Nonetheless, one needs to understand how to consistently halve the degrees of freedom to arrive at the correct physical partition function. In the following Section, we will attempt to recover the partition function trace $Z^{(\delta'', \bar{\delta}'')}_{(\delta',\bar{\delta}')} $ in \eqref{orbifoldZ} in the doubled geometry formalism.

\section{Closed string partition functions of $S^1/\mathbb{Z}_N$ orbifolds}
\label{Sorbifold}

\subsection{Poisson resummations and chiral factorization}

We begin with the action 
$$
S = \frac{\pi}{2}R^2 dX \wedge \ast dX + \frac{\pi}{2R^2} d\tilde{X} \wedge \ast d\tilde{X} + i\pi dX \wedge d\tilde{X}
$$
As discussed earlier, one needs to impose a suitable constraint to recover the physical theory. The natural variables for this constraint are 
\be
P = \frac{1}{\sqrt{2}} \left(  RX + \frac{1}{R}\tilde{X} \right),\qquad Q = \frac{1}{\sqrt{2}} \left( RX - \frac{1}{R}\tilde{X} \right)
\ee
with the constraint being the chiral factorization
\be
\label{constraint}
\partial_z P = \partial_{\bar{z}}Q = 0.
\ee
In terms of $P,Q$,  the action reads
\be
S = \frac{\pi}{2} dP \wedge \ast dP + \frac{\pi}{2} dQ \wedge \ast dQ.
\ee
We now consider the different classical backgrounds differentiated by their cohomology classes as defined on the toroidal worldsheet with cycles $\alpha_1, \alpha_2$. Define
$$
L = dX + \omega, \qquad \omega \in H^1 \left( \sigma, \mathbb{Z} \right)
$$
Imposing the translational twists by phases $\delta$, we can explicitly write $L$ as 
\bea
\label{doubledL}
L &=& dX + (n + \delta' ) \alpha_2 + (m + \delta'') \alpha_1, \\
\tilde{L} &=& d\tilde{X} + (\bar{n} + \bar{\delta}' ) \alpha_2 + (\bar{m} + \bar{\delta}'') \alpha_1.
\eea
In \cite{Berman}, the partition function was shown to be equivalent to the original $S^1$ partition function after keeping the relevant chiral modes by invoking the constraint. Here, we would like to extend the results to $\mathbb{Z}_N$ orbifolds generated by translations in the $S^1$ and dual $S^1$. Roughly speaking, this amounts to studying whether the Poisson resummation steps which were used to reproduce the closed string partition functions are consistent with the addition of the twists. If the doubled formalism is quantum mechanically identical to the original closed string theory, the Poisson resummations should reproduce the phase factors in the shifted lattice sum $Z^{(\delta'', \bar{\delta}'')}_{(\delta',\bar{\delta}')}$ in the closed string theory partition function after chiral factorization. 

The instanton sector of the partition function reads
\bea
\label{startZ}
Z &=& \sum_{\{ n,\bar{n},m \bar{m}  \}} \textrm{exp} \Bigg[  - \left( R(\n)+R^{-1}(\barn) \right)^2 \frac{\pi |\tau|^2}{4\tau_2} \cr
&&+ 
\left( R(\n)+R^{-1}(\barn) \right)\left( R(\m)+R^{-1}(\barm) \right) \frac{\pi \tau_1}{2\tau_2} \cr
&&-\left( R(\m)+R^{-1}(\barm) \right)^2 \frac{\pi}{4\tau_2} \Bigg] \times \textrm{exp}
\Bigg[  - \left( R(\n)-R^{-1}(\barn) \right)^2 \frac{\pi |\tau|^2}{4\tau_2} \cr
&&+  \left( R(\n)-R^{-1}(\barn) \right) \left( R(\m)-R^{-1}(\barm) \right) \frac{\pi \tau_1}{2\tau_2} \cr
&&- \left( R(\m) - R^{-1}(\barm) \right)^2 \frac{\pi }{4\tau_2}  \Bigg],
\eea
where the first factor corresponds to $P$ and the second to $Q$. The phase factors arising
from the topological term read 
\be
\textrm{exp}\left[ i\pi ((\n)(\barm) - (\m)(\barn)) \right]
\ee
In the absence of the twists, the topological term contributes a relative sign to each term in the partition sum. To separate the sum, we write 
$$
R^2 = \frac{p}{q}, \,\,\,p,q\,\,\textrm{coprime integers},\qquad k = pq
$$
and thus, 
\be
 \left( R(\n) \pm R^{-1}(\barn) \right) = \sqrt{k} \left(  \frac{\n}{q} \pm \frac{\barn}{p} \right).
\ee
Substituting 
$n = q(c+\gamma_q), m  = q'(d+\gamma'_q)  $ and $\bar{n} = p(\bar{c} + \gamma_p), \bar{m} = p' (\bar{d} + \gamma'_p)$, 
where 
$c,\bar{c},d,\bar{d} \in \mathbb{Z}$ and $\gamma_k, \gamma'_k \in \left\{ 0, \frac{1}{k},\ldots,\frac{k-1}{k} \right\}$, and letting
$h= c+ \bar{c}, l = c - \bar{c}, i = d+\bar{d}, j = d-\bar{d},$
the partition function then reads (below, the set of summation indices is $I = \{ h,l,i,j,\gamma_\pm, \gamma'_\pm, \phi,\theta  \})$
\bea
\label{s1partitionf}
Z &=& \sum_I \frac{1}{2} \textrm{exp} \left[ 
-\frac{k\pi}{4} \left\{ (h+\gamma_+)^2 \frac{|\tau|^2}{\tau_2} - 2(h+\gamma_+)(i+\gamma'_+)\frac{\tau_1}{\tau_2} + (i+\gamma'_+)^2\frac{1}{\tau_2} \right\} + 2\pi i \left( \phi h + \theta i\right) \right] \cr
&&\times
\frac{1}{2} \textrm{exp} \left[ 
-\frac{k\pi}{4} \left\{ (l+\gamma_-)^2 \frac{|\tau|^2}{\tau_2} - 2(l+\gamma_-)(j+\gamma'_-)\frac{\tau_1}{\tau_2} + (j+\gamma'_-)^2\frac{1}{\tau_2} \right\} - 2\pi i \left( \phi l + \theta j\right) \right] \cr
&&\textrm{exp} \left[ \frac{i\pi k}{2} \left( (l +\gamma_-)(i+\gamma'_+)-(h+\gamma_+)(j+\gamma'_-) \right) \right]
\eea
where we have defined 
\be
\label{gammadef}
\gamma_\pm \equiv \gamma_q +\delta' \pm (\gamma_p + \bar{\delta}'),\qquad 
\gamma'_\pm \equiv \gamma'_q +\delta'' \pm (\gamma'_p + \bar{\delta}''),
\ee
and we have inserted in the factors 
$$
\sum_{\phi,\theta = 0,\frac{1}{2}} \frac{1}{2} \textrm{exp} \left[ 2\pi i (\phi (h-l) + \theta (i - j) ) \right] 
$$
since the terms $h-l$ and $i-j$ are even. Note that each square-bracketed term in \eqref{s1partitionf} corresponds to the contributions from $P$, $Q$ and the topological term respectively. Let us write 
$$
Z= Z_P Z_Q Z_{top}
$$
We now perform a Poisson resummation of $i,j$, replacing them with $r,s$ respectively. After some algebra, this yields (the topological term is absorbed into each of the following pieces)
\bea
Z_P &=& \sum_{J_p} \frac{1}{2} \sqrt{\frac{4\tau_2}{k}} \textrm{exp} \Bigg[ -\frac{k\pi}{4} \left\{ 
(h+\gamma_+)^2 \frac{|\tau|^2}{\tau_2} - 2\gamma_+'(h+\gamma_+)\frac{\tau_1}{\tau_2} 
+ \frac{(\gamma'_+)^2}{\tau_2} \right\} + 2\pi i \phi h  \cr
&&+\frac{i\pi k}{2}(l+\gamma_-)\gamma'_+ -\frac{4\pi \tau_2}{k} \left( r - \theta + ik \frac{(h+\gamma_+)\tau_1}{4\tau_2} - \frac{ik\gamma'_+}{4\tau_2} - \frac{k}{4}(l + \gamma_-) \right)^2 \Bigg] \cr
&=& \sum \sqrt{ \frac{\tau_2}{k} } \textrm{exp} \Bigg[ \tau_2 \left( 
-\frac{k\pi}{4} (h+\gamma_+)^2 - 4\pi k \left(  \frac{r-\theta}{k} - \frac{1}{4}(l+\gamma_-)  \right)^2 \right) \cr 
&&- 2\pi i \tau_1 (h+\gamma_+)(r-\theta-\frac{k}{4}(l-\gamma_-)) + 2\pi i \phi h + 2\pi i (r-\theta)\gamma'_+ \Bigg] 
\eea
and similarly,
\bea
Z_Q &=& \sum_{J_q} \sqrt{\frac{\tau_2}{k}} \textrm{exp} \Bigg[ \tau_2 \left(
-\frac{k\pi}{4} (l+\gamma_-)^2 - 4\pi k \left(  \frac{s-\theta}{k} - \frac{1}{4}(h+\gamma_+)  \right)^2 \right) \cr 
&&+ 2\pi i \tau_1 (l+\gamma_-)(s-\theta-\frac{k}{4}(h+\gamma_+)) - 2\pi i \phi l + 2\pi i (-r +\theta)\gamma'_- \Bigg] 
\eea
where the summation indices are $J_p= 
\{ h,l,r,s,\phi,\theta, \gamma_\pm \gamma'_+ \}, J_q= 
\{ h,l,r,s,\phi,\theta, \gamma_\pm \gamma'_- \} $. At this point, the above expressions are of the same 
form as that presented in \cite{Berman}, with the shift factors contained in the $\gamma_\pm, \gamma'_\pm$ as shown in \eqref{gammadef}. We have used the Poisson summation formula in \eqref{Poisson}. Then, invoking
\be
\tau_2 (a^2 + b^2) - 2i \tau_1 ab = \frac{1}{2i} \left( \tau(a + b)^2 - \bar{\tau} (a- b)^2 \right)
\ee
we can write the partition function as 
\bea
\label{finalpart}
Z &=& \sum  \frac{\tau_2}{k} \textrm{exp} \left[ i\pi k \tau \frac{p^2_L}{2} - i\pi k \bar{\tau} \frac{p^2_R}{2}
+2\pi i (\phi h + (r-\theta) \gamma'_+) \right] \cr
&&\times
\textrm{exp}\left[ i\pi k \tau \frac{q^2_L}{2} - i\pi k \bar{\tau} \frac{q^2_R}{2}
+2\pi i (-\phi l + (s+\theta) \gamma'_-) \right] \times \textrm{exp} \left[ 2\pi i ( w \delta'' + \bar{w} \bar{\delta}''   )    \right] \nonumber \\
\eea
where the left and right-moving momentas are defined as
\be
p_L = \frac{1}{2} (h + \gamma_+) - 2 \left(  \frac{r-\theta}{k} - \frac{1}{4}(l+\gamma_-)     \right),\,\,\,
p_R = \frac{1}{2} (h + \gamma_+) + 2 \left(  \frac{r-\theta}{k} - \frac{1}{4}(l+\gamma_-)     \right)
\ee
\be
q_L = \frac{1}{2} (l + \gamma_-) - 2 \left(  \frac{s+\theta}{k} + \frac{1}{4}(h+\gamma_+)     \right),\,\,\,
q_R = \frac{1}{2} (l + \gamma_-) + 2 \left(  \frac{s+\theta}{k} + \frac{1}{4}(h+\gamma_+)     \right)
\ee
We now sum over all $\gamma'_q, \gamma'_p$. Since
\be
\sum^{N-1}_{u=0} \textrm{exp} (2\pi i \alpha u/N) =  
\begin{aligned}
\begin{cases}
N\,\,\textrm{if}\,\,\alpha\equiv 0 \,\,\textrm{mod}\,\,N\\ 
0\,\,\textrm{otherwise}
\end{cases}
\end{aligned},
\ee
the summation over $\gamma'_q, \gamma'_p$ is reduced to a restricted sum over all $r,s,\theta$ as follows
\be
\label{rs->w}
r+s \equiv 0 \,\,\textrm{mod}\,\,q,\qquad r-s-2\theta \equiv 0\,\,\textrm{mod} \,\,p.
\ee
with a factor of $k=pq$ which cancels the $1/k$ factor in \eqref{finalpart}. From \eqref{rs->w}, we can introduce integers $w, \bar{w} \in \mathbb{Z}$ to write the momenta neatly as 
\bea
p_L &=& \frac{n+\delta'}{q} - \left( \frac{w}{p} + \frac{\bar{w}}{q} \right),\,\cr
p_R &=& \frac{\bar{n}+ \bar{\delta}'}{p} + \left( \frac{w}{p} + \frac{\bar{w}}{q} \right) \cr
q_L &=& -\frac{\bar{n}+\bar{\delta}'}{q} - \left( \frac{w}{p} - \frac{\bar{w}}{q} \right),\,\cr
q_R &=& \frac{n+\delta'}{q} + \left( \frac{w}{p} - \frac{\bar{w}}{q} \right).
\eea
But if we keep track of the phases $\delta', \delta''$, we see that in the momenta to be retained after the chiral projection, we should have both $\bar{\delta}', \delta'$. Comparison with \eqref{orbifoldZ} reveals that since the phases $\delta'', \bar{\delta}''$ couple to $w,\bar{w}$, the latter are the integers that count the effective momenta and winding number modes. Consistency with the original orbifold partition function leads to stringent constraints on what are the appropriates modes to retain when we perform the chiral factorization in accordance with $\partial_{\bar{z}}P = \partial_z Q = 0$.

In \cite{Berman}, the final step in demonstrating the equivalence between the ordinary partition function and that of the doubled CFT is to identify $p_L$ and $q_R$ as the appropriate chiral and anti-chiral pieces. This appears to be compatible with the constraint $\partial_{\bar{z}}P = \partial_z Q = 0$, but we note that in the Poisson resummation process, the topological term does mix modes between $P$ and $Q$, and thus labeling the modes by either $P$ or $Q$ is a subtle procedure which has to be treated with care. 

Consider first a $\mathbb{Z}_N$ shift `symmetric' orbifold which acts equally on the left-and right-moving sectors. In this case, there is no orbifold action on $\tilde{X}$ which corresponds to taking $\bar{\delta}' = \bar{\delta}'' = 0$. The independent modes to be summed over are $n-\bar{w}$ and $w$. Removing the parts which depend on $p_R, q_L$ requires these contributions to be factorized out and this can be done since we can treat $\bar{n} + w$ and $\bar{w}$ as independent modes. After that, the partition function of this 'symmetric' orbifold reads (after relabeling $w \rightarrow -w$)
\be
\label{Zsymmetric}
Z_{sym} = \sum_{n-\bar{w}, w}  e^{-2\pi i \delta'' w } q^{\frac{1}{4} p^2_L }{\bar{q}}^{\frac{1}{4} q^2_R}
\ee
This is indeed the correct expression as we have briefly discussed in Section \ref{Prelim}. Recall that 
\eqref{Zsymmetric} is nothing but $\textrm{Tr}_{\delta'} \left( \hat{s}^{N\delta''} q^{\frac{1}{4} p^2_L } \bar{q}^{\frac{1}{4} q^2_R} \right)$. 

Unfortunately, such an equivalence appears to fail when we consider the case in which the orbifold shift acts asymmetrically, i.e. for non-zero $\bar{\delta}'', \bar{\delta}'$. If we insist on factorizing out $p_R, q_L$, we can immediately see that we cannot capture the effect of $\bar{\delta}'$. However, the factor $e^{2\pi i \bar{\delta}'' \bar{w}}$ is in the right place though, since the insertion of the dual translation operator couples to the winding modes (whereas as we have seen, the insertion of the translation (in $X$) operator couples to the momentum modes). Of course, this ought to be further augmented by multiplying a phase factor $e^{-2\pi i \delta'' n}$ since it is $n-\bar{w}$ that is present in the summation as winding numbers. 

One may attempt to remedy this situation by studying if it's possible to express the squares of sum in the exponential arguments of the various momenta differently, such that we retain both $\delta', \bar{\delta}'$. For example, suppose we redefine the left- and right-moving degrees of freedom as
\bea
\label{momentaredefin}
\tilde{p}_L &=& \frac{1}{\sqrt{2}} \left(\frac{\n}{q} + \frac{\barn}{p} -2\frac{\bar{w}}{q} \right) \cr
\tilde{p}_R &=& \frac{1}{\sqrt{2}} \left(\frac{\n}{q} + \frac{\barn}{p} +2\frac{w}{q} \right) \cr
\tilde{q}_L &=& \frac{1}{\sqrt{2}} \left(\frac{\n}{q} - \frac{\barn}{p} -2\frac{w}{q} \right) \cr
\tilde{q}_R &=& \frac{1}{\sqrt{2}} \left(\frac{\n}{q} - \frac{\barn}{p} -2\frac{\bar{w}}{q} \right),
\eea
and we attempt to retain the relevant parts of the partition function by some form of chiral factorization, the independent modes to sum over can only be $\{ \bar{n} + 2w,n\}$ and $\{\bar{n}, n - 2\bar{w} \}$. These two sets, unlike the earlier case of the symmetric orbifold, are not independent, but we can still separate them into chiral blocks by summing over even and odd $n,\bar{n}$ separately. Let's define 
$u = \bar{n} + 2w,v = n - 2\bar{w}$. Then the parity of $u,v$ follows that of $n,\bar{n}$ respectively. 
Thus the complete partition function is a sum of four distinct sectors depending on whether $n,\bar{n}$ are even or odd, i.e.
$$
Z = \sum_{n,\bar{n} \in 2\mathbb{Z}, 2\mathbb{Z}+1}
Z^{(\tilde{p}_L)}_{v+\bar{n}}   Z^{(\tilde{q}_R)}_{v-\bar{n}} Z^{(\tilde{q}_L)}_{n-u} Z^{(\tilde{p}_R)}_{n+u}. 
$$
The chiral factorization procedure then amounts to setting $Z^{(\tilde{q}_L)} Z^{(\tilde{p}_R)}=1$. Yet this is problematic because the phase factors $\delta'', \bar{\delta}''$ must couple to $w,\bar{w}$ with the latter being momentum and winding integral modes. This clearly does not agree with the expressions in \eqref{momentaredefin}. 

Thus far, we have followed rather closely the approach in \cite{Berman}. When the orbifold action acts purely on $X$, and is thus symmetric and geometric, the closed string partition function can be recovered in a consistent fashion by chiral factorization. When the orbifold action is asymmetric and thus non-geometric, we find that the same procedure seems to require some modification. It turns out that we need to couple additional phase factors to the partition trace when $\bar{\delta}', \bar{\delta}'' \neq 0$. As we shall shortly point out, exactly the same phase factors arise in the conventional theory too. 

\subsection{Some pairing phase factors}
 
We have argued that it does not seem possible to reduce the instanton part of the doubled partition function to the correct physical one by merely assuming $d\tilde{X} = (\bar{n} + \bar{\delta}')\alpha_2 + (\bar{m} + \bar{\delta}'')\alpha_1 $. Let us introduce some additional phase factors when computing $Z^{(\delta'', \bar{\delta}'')}_{(\delta',\bar{\delta}')}$ right from the outset. These phase factors measure the asymmetric (and hence non-geometric) nature of the orbifold. We now compute the doubled partition function with 
the inclusion of the following phase factor
\be
\label{pairingphasefactors}
\textrm{exp} \left[ 2\pi i \left( \bar{\delta}'m - \bar{\delta}''(n+\delta') \right)        \right]
\ee
in the sum over all modes in the starting step \eqref{startZ}. We find that we can indeed recover the appropriate doubled lattice sum reducible to \eqref{orbifoldZ} after chiral factorization. Since we have already illustrated in detail this procedure (a rather simple extension of what was done in \cite{Berman}), we choose not to be repetitive and instead present a somewhat shorter derivation that leads to the same partition function. We will perform separate Poisson resummations $m \rightarrow w,\bar{m} \rightarrow \bar{w}$ which are the winding modes along the Euclidean time direction. It turns out that after resumming, the Lagrangian admits a unique splitting into a sum of the square of four distinct momenta terms, two of which are multiplied to $\tau$ and the other two to $\bar{\tau}$, thus allowing us to interpret them as $p_{L,R}, q_{L,R}$. Although this shorter derivation is presented mainly for the purpose of efficiency, we should bear in mind that for a slightly more rigorous argument that allows us to discard half of the chiral modes away in the final step, the longer derivation presented earlier is tighter in logic. In the next section on more general toroidal backgrounds, we shall follow the more expedient derivation, leaving the longer exercise for any keen reader. 

In any case, the fact that the doubled partition function can be chirally factorized is not trivial. The inclusion of the topological term plays a crucial role. Its dependence on the winding numbers exactly accounts for its needed appearance to ensure that the doubled sigma model is invariant under large gauge transformations. On the other hand, in our computations, this topological term governs the correct mixture of the momenta and winding degrees of freedom of $X$ and $\tilde{X}$. 

Including the phase factors \eqref{pairingphasefactors}, we thus begin with the instanton sum
\bea
\label{withphases}
Z_{cl.} &=& \sum_{n,m,\bar{n},\bar{m}} \textrm{exp}\left[ -2\pi i (\bar{\delta}'' (n+\delta') +\delta'' \bar{\delta}' ) -\frac{\pi |\tau|^2}{2\tau_2} \left( R^2(n+\delta')^2 + R^{-2}(\bar{n}+ \bar{\delta}')^2  \right) \right] \cr
&&\times \textrm{exp} \left[ -\frac{\pi R^2}{2\tau_2}(m+\delta'')^2 +
2\pi i \left( \frac{R\tau_1}{2i\tau_2}(n+\delta') - \frac{1}{2R}(\bar{n} - \bar{\delta}') \right)R(m+\delta'') \right]\cr
&&\times \textrm{exp} \left[ -\frac{\pi}{2\tau_2R^2}(\bar{m}+\bar{\delta}'')^2 +
2\pi i \left( \frac{\tau_1}{2i R \tau_2}(\bar{n}+\bar{\delta}') + \frac{R}{2}(n + \delta') \right)R^{-1}(\bar{m}+\bar{\delta}'') \right] \nonumber\\
\eea
Then, we proceed to perform separate Poisson resummations $m \rightarrow w$ and $\bar{m} \rightarrow{w}$ starting from \eqref{withphases}. After some algebra, we arrive at\footnote{Each of the Poisson resummation process generates $\sqrt{2\tau_2/R^2}$ and $\sqrt{2\tau_2 R^2}$ which combine to yield a pre-factor $2\tau_2$ in \eqref{withphases}. The absence of $R$ holds some implications for higher-genus worldsheets and we shall discuss this later in Section \ref{Genus}.}
\bea
\label{resummedaction}
Z_{cl.} &=& \sum_{n,w,\bar{n},\bar{w}} (2\tau_2) \textrm{exp} \Bigg[ -2\pi \tau_2 \left( \frac{w}{R} + \frac{\tau_1 R}{2i\tau_2}(n+\delta') - \frac{1}{2R}(\bar{n} - \bar{\delta}') + i\frac{\delta'' R}{2\tau_2}    \right)^2 -\frac{\pi \delta''^2 R^2}{2\tau_2}      \cr
&& -2\pi \tau_2 \left( \frac{\bar{w}}{R} + \frac{\tau_1}{2i\tau_2 R}(\bar{n}+\bar{\delta}') + \frac{R}{2}(n + \delta') + i\frac{\bar{\delta}''}{2\tau_2 R}    \right)^2 -\frac{\pi \bar{\delta}''^2}{2\tau_2 R^2} \cr
&& + 2\pi i \left(\frac{\tau_1}{2i\tau_2 R^2}(\bar{n}+\bar{\delta}') + \frac{1}{2}(n +\delta') \right) \bar{\delta}'' 
+2\pi i \left( \frac{\tau_1 R^2}{2i\tau_2}(n+\delta') - \frac{1}{2}(\bar{n} - \bar{\delta}') \right) \delta'' \cr
&&-\frac{\pi |\tau|^2}{2\tau_2} \left( R^2 (n+\delta')^2 +R^{-2} (\bar{n}+ \bar{\delta}')^2  \right) -2\pi i (\bar{\delta}''n + \delta' \bar{\delta}'') \Bigg]
\eea
Consider first the terms which are multiplied by the phase $\delta'', \bar{\delta}''$. They amount to the phase factors 
\be
\label{pairingphase}
\textrm{exp}\left[- 2\pi i \delta'' (w+ \bar{\delta}') - 2\pi i \bar{\delta}''(\bar{w} + n +\delta' ) \right]
\ee
which turns out to be the appropriate phase factors that should appear in the asymmetric orbifold partition function. The other terms in the exponential argument simplify to read 
\bea
&&\frac{i\pi \tau}{2} \left( \left[ (n+\bar{w}+\delta')R + \frac{(w+\bar{\delta}')}{R} \right]^2 + \left[ \frac{\bar{n} - w}{R}+ \bar{w}R \right]^2 \right) \cr
&&-\frac{i\pi \bar{\tau}}{2} \left( \left[ (n+\bar{w}+\delta')R - \frac{(w+\bar{\delta}')}{R} \right]^2 + \left[ \frac{\bar{n} - w}{R} - \bar{w}R \right]^2 \right)
\eea
Defining 
\bea
&&p_L =  (n+\bar{w}+\delta')R + \frac{(w+\bar{\delta}')}{R} ,\,\,
p_R = \frac{\bar{n} - w}{R}+ \bar{w}R  \cr
&&q_L =\frac{\bar{n} - w}{R} - \bar{w}R,\,\,
q_R = (n+\bar{w}+\delta')R - \frac{(w+\bar{\delta}')}{R},
\eea
we see that the doubled partition function, including the quantum fluctuations (see Section \ref{Quantum}), can be simplified to read
\be
\label{finalorbifoldZ}
Z = \frac{1}{|\eta|^2} \sum_{n,\bar{n},w,\bar{w}} 
e^{- 2\pi i \left[ \delta'' (w + \bar{\delta}' )+ \bar{\delta}''(\bar{w} + n + \delta' ) \right]} q^{\frac{1}{4}(p^2_L +q^2_L)} \bar{q}^{\frac{1}{4}(p^2_R +q^2_R)}.
\ee
Factoring out $p_R$ and $q_L$, the complete partition function now agrees with what we would expect in a conventional computation. It is noteworthy that the discarded modes do not capture any of the orbifold phases $\{\delta', \bar{\delta}', \delta'', \bar{\delta}''\}$. We view this as a simple consistency check of the chiral factorization procedure. Our result demonstrates that at the asymmetric orbifold point, one can consistently remove the unphysical states from the doubled theory to recover the spectrum of the asymmetric orbifold. The non-geometric action of the orbifold manifests itself as complex phase factors when one writes down the lattice sum. 

Our introduction of the phase factors \eqref{pairingphasefactors} might appear to be rather contrived, hence we must now point out that these factors actually arise in an identical fashion in the ordinary theory. Beginning from \eqref{orbifoldZ}, one can check that a Poisson resummation of $w$ yields the instanton sum
\be
\label{PorbifoldZ}
Z_{cl.} \sim \sum_{m,n} e^{\left[ 2\pi i \left( \bar{\delta}'m - \bar{\delta}''(n+\delta') \right) \right]} e^{-\frac{\pi R^2}{\tau_2}  \lvert \tau(n + \delta') - (m + \delta'') \rvert^2    }
\ee
where we observe that the complex phase factors that arise are precisely \eqref{pairingphasefactors}. It was noted in \cite{Aoki} that generically, the partition functions of asymmetric orbifolds can be constructed from the chiral blocks of associated symmetric orbifolds where the orbifold action $\mathcal{G}_{L,R}$ is symmetrized in each block $\mathcal{F}$ and that sewing these blocks together requires, in general for modular covariance, a pairing matrix $K$. 

Symbolically, as explained in \cite{Aoki}, we can write the `block' $Z^g_h$ of the asymmetric orbifold with $h=(h_L;h_R), g=(g_L;g_R)$ as
\be
Z^g_h (p_L, p_R;\tau,\bar{\tau}) = \mathcal{F}^{g_L}_{h_L} (p_L;\tau) \overline{\mathcal{F}^{g_R}_{h_R} (p_R;\tau)}
\ee
with the full partition function being 
\be
Z_G (\tau,\bar{\tau}) = \frac{1}{|G|} \sum_{g,h} \sum_{p_{L,R}} K(p_{L},p_{R},h_L,h_R,g_L,g_R)  \mathcal{F}^{g_L}_{h_L} (p_L;\tau) \overline{\mathcal{F}^{g_R}_{h_R} (p_R;\tau)}
\ee
where $|G|$ is the order of the orbifold group.
It is interesting to see that the doubled sigma model presents a theory for which the computation of the one-loop partition function involves naturally a pairing of some left and right chiral blocks. In this picture, the phase factors in \eqref{pairingphase} appears as a pairing phase matrix that pairs up the blocks. It would be interesting to explore this connnection genuinely by working in the operator formalism \cite{future}. What we have demonstrated is the quantum equivalence between the doubled sigma model and the conventional one via a simple chiral factorization process. In the following Sections, we shall continue to extend our result  in several natural directions: (i)the general toroidal background with $B$ field (ii)higher-genus worldsheet (iii)$\mathcal{N}=1$ worldsheet supersymmetry.

\section{More general toroidal compactifications}
\label{Toroidal}

Now, we lift our analysis of the previous two sections to higher-dimensional toroidal backgrounds with the general metric $G_{mn}$ and the Kalb-Ramond field $B_{mn}$. Before that, let us first derive the lattice sum of such a background (without orbifolding) by resumming the winding modes along the Euclidean time direction. 
In the following treatment, all $X^i$ are dimensionless periodic coordinates coordinates. The $d-$dimensional torus is $T^d = \mathbb{R}^d / \Lambda^d$ where $\Lambda$ is the compactification lattice with metric $G_{mn}$. Similar to \eqref{doubledL}, we can write
\bea
\label{TdoubledL}
L^r &=& dX^r + n^r \alpha_2 +  m^r  \alpha_1, \\
\tilde{L}_r &=& d\tilde{X}_r + \bar{n}_r \alpha_2 + \bar{m}_r \alpha_1,
\eea
with the choice of the position of the index $r$ being such that all winding numbers are integral in \eqref{TdoubledL}. The classical action reads
\bea
\label{toroidalaction}
-S_{cl} &=& \frac{\pi}{ \tau_2} G_{mn} \left( w^m_2 - w^m_1 \tau\right) \left(w^n_2 - w^n_1 \bar{\tau} \right) + 2\pi i B_{mn} w^m_1 w^n_2 \cr
&=& - \left(\pi w^m_2 A_{mn} w^n_2 - 2\pi i R_m w^m_2 + \frac{\pi G_{mn}}{\tau_2} w^m_1 w^n_1 |\tau|^2\right)
\eea
where we have defined 
$$
A_{mn} = \frac{G_{mn}}{\tau_2},\,\,\, R_m = B_{mn}w^n_1 - i \frac{G_{mn} w^n_1 \tau_1}{\tau_2}
$$
A Poisson resummation of all $w^m_2 \rightarrow n^m$ yields 
\bea
\label{toroidalactionresum}
-S_{cl} &=& - \pi \left( n_m - R_m + i \frac{\tau_1 w_m}{\tau_2} \right) \tau_2 G^{mn}
\left(  n_n - R_n + i \frac{\tau_1 w_n}{\tau_2} \right) - \frac{w^2 |\tau|^2}{\tau_2} \cr
&=& -\pi \left( \tau_2 \left( n_a G^{ab} n_b + w_a G^{ab} w_b - 2B_{ab} w^b n^a + w^m w^n B_{im}B^i_n \right) -2i \tau_1 w^a n_a \right)
\eea
It turns out that the above sum takes a more illuminating form
\be
-S_{cl}= \frac{\pi i \tau P_L^2}{2} - \frac{\pi i \bar{\tau} P_R^2}{2}
\ee
where the  left- and right-moving momenta depend on the fields and the vielbeins
$$
P_{\{L,R\}}a = e^m_a \left( w_m \pm (G_{mk}n^k + B_{mn} w^n) \right),\qquad G^{mn} \equiv \eta^{ab} e^m_a e^n_b
$$
Recall that the factorized T-duality element that relates $\tilde{X}$ to $X$ induces the following action on the metric and the $B$ field. 
\be
\label{actionOnE}
\tilde{G}^{mn} = \left( G_{mn} - B_{ma} G^{ab} B_{bn} \right)^{-1},\,\,\,
\tilde{B}^{mn} = \left(  B_{mn} - G_{ma} \left(B^{-1}\right)^{ab} G_{bn} \right)^{-1}
\ee
Yet we do not need explicit knowledge of \eqref{actionOnE} since we are starting from \eqref{PQaction}. 
We now proceed to compute the partition function of the doubled sigma model and investigate whether it is possible to discard the unphysical modes by chiral factorization. Our starting point is the action 
\eqref{PQaction}, with the topological term $i\pi \,dX \wedge d\tilde{X}$.  Substituting \eqref{PQdefinition} into \eqref{PQaction}, we find that the various terms assemble to be
\bea
\label{PQactionexpanded}
Z_{cl.} &=& \sum_{\{ n,\bar{n},m,\bar{m}  \}} \textrm{exp}\Bigg[ -\frac{\pi |\tau|^2}{4\tau_2} \left[ \left( e^a_i - B_{im}e^{ma} \right)n^i + e^{ja} \bar{n}_j    \right]^2 
-\frac{\pi}{4\tau_2}\left[ \left( e^a_i - B_{im}e^{ma} \right)m^i + e^{ja} \bar{m}_j    \right]^2 \cr
&&+\frac{\pi \tau_1}{2\tau_2} \left[ \left( e^a_i - B_{im}e^{ma} \right)n^i + e^{ja} \bar{n}_j    \right]
\left[ \left( e^a_i - B_{im}e^{ma} \right)m^i + e^{ja} \bar{m}_j    \right] \cr
&-&\frac{\pi |\tau|^2}{4\tau_2} \left[ \left( e^a_i + B_{im}e^{ma} \right)n^i - e^{ja} \bar{n}_j    \right]^2 
-\frac{\pi}{4\tau_2}\left[ \left( e^a_i + B_{im}e^{ma} \right)m^i - e^{ja} \bar{m}_j    \right]^2 \cr
&&+\frac{\pi \tau_1}{2\tau_2} \left[ \left( e^a_i + B_{im}e^{ma} \right)n^i - e^{ja} \bar{n}_j    \right]
\left[ \left( e^a_i + B_{im}e^{ma} \right)m^i - e^{ja} \bar{m}_j    \right] \cr
&&+i\pi \left( n^j \bar{m}_j - \bar{n}_k m^k      \right) \Bigg]
\eea
We  then proceed similarly to the $S^1$ target space case. To simplify the algorithm, we shall perform Poisson resummation of $m$ and $\bar{m}$, replacing them with $w$ and $\bar{w}$. As mentioned earlier, this effectively turns out to be equivalent to resumming $m\pm \bar{m}$ which was done for the $S^1$ case earlier. Let us first check the result in the simpler case of vanishing $B$ field. One can see that when $B=0$, the terms in $m$ and $\bar{m}$ do not mix, and the Poisson resummation can be done independently to yield 
\bea
\label{withoutBfield}
Z_{cl} &=& (2\tau_2)^d \sum_{\{ n,\bar{n},w,\bar{w}  \}} \textrm{exp}\Bigg[ 
-2\pi \tau_2 G_{ij}  \left( \bar{w}^i + \frac{1}{2}n^i -\frac{i\tau_1}{2\tau_2}G^{ki} \bar{n}_k \right)
\left( \bar{w}^j + \frac{1}{2}n^j -\frac{i\tau_1}{2\tau_2}G^{kj} \bar{n}_k \right) \cr
&&-2\pi \tau_2 G^{ij}  \left( w_i + \frac{1}{2}\bar{n}_i +\frac{i\tau_1}{2\tau_2}G_{ki} n^k \right)
\left( w_j + \frac{1}{2}\bar{n}_j + \frac{i\tau_1}{2\tau_2}G_{kj} n^k \right)\cr
&&-\frac{\pi |\tau|^2 }{2\tau_2} \left( G_{ij} n^i n^j + G^{ik} \bar{n}_i \bar{n}_k \right) \Bigg] \cr
&=&(2\tau_2)^d
\sum_{\{ n,\bar{n},w,\bar{w}  \}} \textrm{exp} \Bigg[ 
 -2\pi \tau_2  \Bigg( G_{ij} (\bar{w}^i + \frac{1}{2}n^i )(\bar{w}^j + \frac{1}{2}n^j)
+G^{ij} (w_i + \frac{1}{2}\bar{n}_i )(w_j + \frac{1}{2} \bar{n}_j)\cr
&&+\frac{1}{4} (G_{ij} n^i n^j + G^{ij} \bar{n}_i \bar{n}_j ) \Bigg) +2\pi i \tau_1 \left(  \bar{n}_k \bar{w}^k - n^k w_k     \right) \Bigg] \cr
&=& (2\tau_2)^d \sum_{\{ n,\bar{n},w,\bar{w}  \}} \textrm{exp}\Bigg[ -2\pi \tau_2 \left( G_{ij} \bar{w}^j \bar{w}^i + G^{ij}w_i w_j + \frac{1}{2}(  G_{ij} n^i n^j + G^{ij} \bar{n}_i \bar{n}_j) + G^{ij} \bar{n}_i w_j \right) \cr
&&+ 2\pi i \tau_1 \left( \bar{n}_i \bar{w}^i - n^i w_i \right) \Bigg]. 
\eea
The last line of \eqref{withoutBfield} is identical in form to what we have for the doubled sigma model for a circle target space, with the metric $G_{ij}$ contracting the various indices. After factoring out the appropriate modes, and combining with the quantum part which comes from the oscillators, we again have the correct contribution to the partition function which reads
\be
Z = \frac{1}{|\eta|^{2d}} \sum_{P_{L}, P_{R}} e^{\frac{i\pi}{2} \tau P^2_L}  e^{-\frac{i\pi}{2} \bar{\tau} P^2_R} ,\qquad
P_{(L,R)a} = e^m_a \left( w_m \pm G_{mk} (n^k + \bar{w}^k )\right), \,\,\,P^2 \equiv \eta^{ab} P_a P_b
\ee
We now include the $B$ field. In \eqref{PQactionexpanded}, we can see that it mixes the $m,\bar{m}$ terms.  We find it more convenient to perform a Poisson resummation of $\bar{m} \rightarrow \bar{w}$ first. To see this, let's assemble the $\bar{m}$ terms which read
$$
Z_{cl.} = (2\tau_2)^d\sum_{\{ n,\bar{n},m,\bar{m}  \}} \textrm{exp} \Bigg[  -\frac{\pi}{2\tau_2} G^{ik} \bar{m}_i \bar{m}_k + \left[ \frac{\pi}{\tau_2} B_{ij} G^{jk} m^i + \frac{\pi \tau_1}{\tau_2} (e^m \bar{n}_m - B_{ij} e^j n^i)e^k + i\pi n^k \right] \bar{m}_k + \ldots \Bigg]
$$
After Poisson resumming $\bar{m}$, the terms quadratic in $m$ read $-\frac{\pi}{2\tau_2} m^k m^l B_{kj} G^{jm}B_{ml}$ which cancels away an identical term coming from the $P$ part of the action to yield the term $-\frac{\pi}{2\tau_2} G_{ij} m^i m^j + \ldots.$ in the argument of the exponential. Since the metric can be inverted explicitly, we can proceed to Poisson resum $m$ easily. One can check that if we choose to Poisson resum $m$ first, the quadratic term in $m$ does not allow one to invert the matrix as neatly. The Poisson resummation of $\bar{m}$ yields 
\bea
Z_{cl.} &=& (2\tau_2)^d/2 \sqrt{\textrm{Det} (G)} \sum_{\{ n,\bar{n},m,\bar{w}  \}} \textrm{exp} \Bigg[ 
-\frac{\pi}{2\tau_2} G_{ik} m^i m^k + 2\pi i \Bigg\{ B_{ik} \left(\bar{w}^k + D^k \right) - \frac{1}{2}\bar{n}_i  \cr
&&+ \frac{\tau_1}{4i\tau_2} \left[  (e^-_l e^-_i + e^+_l e^+_i ) n^l + (e^k e^-_i - e^k e^+_i ) \bar{n}_k \right] \Bigg\} m^i  \cr
&&-2\pi \tau_2 G_{ij} (\bar{w}^i + D^i )(\bar{w}^j + D^j) - \frac{\pi |\tau|^2}{4\tau_2} 
\Bigg[ (e^-_i n^i + e^j \bar{n}_j)^2 + (e^+_i n^i - e^j \bar{n}_j)^2 \Bigg] \Bigg] \nonumber \\
\eea
where we have defined 
$$
D^k = \frac{\tau_1}{2i\tau_2} \left(  G^{rk} \bar{n}_r - B_{rm} G^{mk} n^r \right) + \frac{1}{2}n^k,\qquad
e^{\pm}_i = e_i \pm B_{ij} e^j
$$
The term linear in $\bar{m}$ simplify to read 
$$
K_i= B_{ik} (\bar{w}^k + \frac{1}{2}n^k ) - \frac{1}{2} \bar{n}_i + \frac{\tau_1}{2i\tau_2} G_{li} n^l.
$$
After Poisson resumming $m\rightarrow w$, the instanton partition function can be simplified to become
\bea
\label{eventualL}
Z_{cl.} &=&(2\tau_2)^d \sum_{\{ n,\bar{n},w,\bar{w}  \}} \textrm{exp} \Bigg[       -2\pi \tau_2 G^{ik} (w_i + K_i)(w_k + K_k) - 2\pi \tau_2 G_{ij} (\bar{w}^i + D^i)(\bar{w}^j + D^j) \cr
&&-\frac{\pi |\tau|^2}{4\tau_2} \left(  2(n_a n^a + B_{am} G^{mk} B_{bk} n^a n^b ) + 2\bar{n}^a \bar{n}_a - 4 B_{im}G^{mk} n^i \bar{n}_k \right) \Bigg].
\eea
The terms in $\tau_1^2$ can be checked to cancel nicely, and it turns out that 
\eqref{eventualL} can be massaged into the form 
\bea
\label{Lfinal}
Z_{cl.}&=&(2\tau_2)^d \sum_{\{ n,\bar{n},w,\bar{w}  \}} \textrm{exp} \Bigg[ -2\pi \tau_2 \Bigg( G^{ik} w_i w_k + G_{ik}\bar{w}^i \bar{w}^k + \frac{1}{2} G_{ik} n^i n^k + \frac{1}{2} G^{ik} \bar{n}_i \bar{n}_k \cr
&& - G^{ik} \bar{n}_i w_k + G_{ik} n^i \bar{w}^k
+ 2 w_k \bar{w}^n {B^k}_n + w_i n^j {B^i}_j - \bar{w}^i \bar{n}_j {B^j}_i \cr
&&+ \left[ 2\bar{w}^m \bar{w}^n + 2n^m \bar{w}^n + n^m n^n  \right]{B^k}_m B_{kn} \Bigg) +i\pi \tau_1
\left( 2n^i w_i + 2\bar{n}_i\bar{w}^i \right) \Bigg]
\eea
More suggestively terms of $\tau, \bar{\tau}$, \eqref{Lfinal} reads 
\bea
\label{Tfinal}
Z_{cl.}&=&(2\tau_2)^d \sum_{\{ n,\bar{n},w,\bar{w}  \}} \textrm{exp} \Bigg[
\frac{i\pi}{2}  \tau \left( (n+\bar{w} + B(n+\bar{w}) + w)^2 + (\bar{n} - w -B\bar{w} + \bar{w})^2 \right) \cr
&&-\frac{i\pi}{2} \bar{\tau} \left( (\bar{n}-w-B\bar{w} - \bar{w})^2 + (-(n+\bar{w}) + B(n+\bar{w}) + w)^2 \right) \Bigg]
\eea
where we have suppressed all indices (raised/lowered by metric $G$) to unveil the resemblance of the structure of the expression to that of the $S^1$ case. As was the case previously, we note that the splitting of the terms into sum of two squares is unique. Recall that in our presentation for the $S^1$ target space, apart from a small ambiguity that arises from the topological term, each term in the doubled sigma model's action can be attributed to either $P$ or $Q$ neatly. This is as it should formally be. Although we have presented a shorter derivation, the final result is identical. The only caveat lies in identifying which terms belong to that of $P$ or $Q$ degrees of freedom.  The expression in \eqref{Lfinal} can be seen to be the correct generalization of the result first presented in \cite{Berman} for the $S^1$ target space. Just as in that scenario, we should identify the independent momentum and winding modes as
\bea
&&\textrm{winding modes}: u\equiv n+\bar{w}, \bar{w} \cr
&&\textrm{momentum modes}: w, v\equiv \bar{n}-w
\eea
Finally, restoring the indices, we can then identify all the left- and right-moving momenta as
\bea
\label{Tmomenta}
p_{La} &=& e^m_a \left( w_m + B_{mk} u^k + G_{mk}u^k \right) \cr
p_{Ra} &=& e^m_a \left( v_m - B_{mk} \bar{w}^k - G_{mk} \bar{w}^k \right) \cr
q_{La} &=& e^m_a \left( v_m - B_{mk} \bar{w}^k + G_{mk} \bar{w}^k \right) \cr
q_{Ra} &=& e^m_a \left( w_m + B_{mk} u^k - G_{mk}u^k   \right) 
\eea
and taking into account the quantum part due to the oscillators, the doubled partition function reads
\be
Z= \frac{1}{|\eta|^{2d}} \sum_{p_{L,R}, q_{L,R}} \textrm{exp} \left[  \frac{1}{2} i\pi \tau p^2_L- \frac{1}{2} i\pi \bar{\tau} p^2_R \right] 
\textrm{exp} \left[ \frac{1}{2}i\pi \tau q^2_L - \frac{1}{2}i\pi \bar{\tau} q^2_R \right] 
\ee
The constraint can then be implemented immediately by discarding the $q_L, p_R$ modes, similar to the $S^1$ case. We have thus shown that starting from the doubled sigma model action, we can arrive at the correct physical theory by chiral factorization. This generalizes the result in \cite{Berman} to general toroidal backgrounds. In several aspects, this is not a trivial consistency check because this derivation is sensitive to how the $B$ field couples to the original $X$ and $\tilde{X}$ terms. The topological term plays an essential role here as well. That we recover the correct partition function at the end is a non-trivial evidence for the structure of the doubled sigma model's Lagrangian and whether it allows us to implement the constraint, at least in this specific computation, as a form of chiral factorization. 

Translational orbifolds can be defined via shifts in the torus lattice, similar to what was discussed in \ref{Sorbifold}. Proving the quantum equivalence is similar in spirit to the $S^1$ case, so we shall leave the exercise to the reader.

\section{Higher-Genus worldsheets}
\label{Genus}

Let $g$ denote the worldsheet genus. Then the complete string partition function reads 
\be
\label{higherG}
Z_g = \int dh\,dX \textrm{exp}\left[ - \frac{1}{4\pi} \int_{M_g} d\tau d\sigma
\sqrt{h} h^{mn} G_{ij} \partial_m X^i \partial_n X^j + \epsilon^{mn} B_{ij} \partial_m X^i \partial_n X^j  \right]
\ee
where we integrate over all worldsheet metrics $h$ that are compatible with genus $g$. Let us define the canonical homological cycles $(a_\alpha, b_\alpha)$ of $M_g$ as follows. Define $\omega_\alpha, \alpha=1,2,\ldots g$ be the holomorphic one-forms that span 
$H_{(1,0)} (M_g, \mathbb{Z})$, and the $g\times g$ period matrix $\tau$ be
\be
\label{periodmatrix}
\tau_{\alpha \beta} = \int_{a_\alpha} \omega_\beta,\qquad \int_{b_\alpha} \omega_\beta = \delta_{\alpha \beta}, \qquad \int_{a_\alpha} dX^i = 2\pi n^i_\alpha, \qquad \int_{b_\alpha} dX^i = 2\pi m^i_\alpha
\ee
where we have defined $n^i_\alpha, m^i_\alpha$ to be the $d\times g-$dimensional winding and momentum vector modes. The complete partition function in \eqref{higherG} can be written as an integral over all the $3g-3$ modular parameters. In the following, we shall first consider the classical instanton sector of the partition function, reserving the discussion of the quantum part to Section \ref{Quantum}. The winding numbers along each cycle are now $d\times g$-dimensional vectors. The partition function reads 
\bea
Z_{cl.}(G,B,\tau) &=& \sum_{n,m}
\textrm{exp} \Bigg[ -\pi
m^{i\alpha} (\frac{1}{\tau_2})_{\alpha \beta} G_{ij} m^{j\beta} -\pi 
n^{i\alpha} (\tau_2 + \tau_1 \frac{1}{\tau_2} \tau_1)_{\alpha \beta} G_{ij} n^{j\beta} \cr
&&+\pi 2n^{i\alpha} (\tau_1 \frac{1}{\tau_2} )_{\alpha \beta} G_{ij} m^{j\beta} + 2i \pi m^{i\alpha} B_{ij} n^{j\alpha} \Bigg]
\eea
This can be Poisson resummed just as in the torus case. The result is known (see for example \cite{Giveon}) and the classical zero modes' part can be cast into the familiar form
\be
\label{Gpartitionfunction}
\textrm{Det}(\tau_2)^{d/2} \sum_{(\vec{k},\vec{m})\in \mathbb{Z}^{2dg}}
q^{\frac{1}{4} P^a_{L\alpha} \tau_{\alpha \beta} P^a_{L\beta}  }\bar{q}^{\frac{1}{4} P^a_{R\alpha} \bar{\tau}_{\alpha \beta} P^a_{R\beta} }
\ee
where the dependence of $p_{L,R}$ on the winding and momentum vectors  $(\vec{k},\vec{m})$ and the fields $G,B$ are identical in form to \eqref{Tmomenta}.

Let us now recover \eqref{Gpartitionfunction} in the doubled formalism. In the following, to avoid cluttering
of indices, we shall suppress the Greek indices, bearing in mind the matrix nature of $\tau$. The various terms assemble to be a straightforward generalization of \eqref{PQactionexpanded}. 
\bea
\label{PQactionexpanded2}
Z_{cl.} &=& \sum_{n, \bar{n}, m, \bar{m}}
\textrm{exp} \Bigg[
 -\frac{\pi}{4} \left[ \left( e^a_i - B_{im}e^{ma} \right)n^i + e^{ja} \bar{n}_j    \right] 
\left(  \tau_2 + \tau_1 \frac{1}{\tau_2} \tau_1  \right) \left[ \left( e^a_i - B_{im}e^{ma} \right)n^i + e^{ja} \bar{n}_j    \right] \cr
&&-\frac{\pi}{4} \left[ \left( e^a_i - B_{im}e^{ma} \right)m^i + e^{ja} \bar{m}_j    \right] \frac{1}{\tau_2} 
\left[ \left( e^a_i - B_{im}e^{ma} \right)m^i + e^{ja} \bar{m}_j    \right]\cr
&&+\frac{\pi }{2} \left[ \left( e^a_i - B_{im}e^{ma} \right)n^i + e^{ja} \bar{n}_j    \right]
\tau_1 \frac{1}{\tau_2}  \left[ \left( e^a_i - B_{im}e^{ma} \right)m^i + e^{ja} \bar{m}_j    \right] \cr
&&-\frac{\pi}{4} \left[ \left( e^a_i + B_{im}e^{ma} \right)n^i - e^{ja} \bar{n}_j    \right] 
\left(  \tau_2 + \tau_1 \frac{1}{\tau_2} \tau_1  \right) \left[ \left( e^a_i + B_{im}e^{ma} \right)n^i - e^{ja} \bar{n}_j    \right] \cr
&&-\frac{\pi}{4} \left[ \left( e^a_i + B_{im}e^{ma} \right)m^i - e^{ja} \bar{m}_j    \right] \frac{1}{\tau_2} 
\left[ \left( e^a_i + B_{im}e^{ma} \right)m^i - e^{ja} \bar{m}_j    \right]\cr
&&+\frac{\pi }{2} \left[ \left( e^a_i + B_{im}e^{ma} \right)n^i - e^{ja} \bar{n}_j    \right]
\tau_1 \frac{1}{\tau_2}  \left[ \left( e^a_i + B_{im}e^{ma} \right)m^i - e^{ja} \bar{m}_j    \right] \cr
&&+i\pi \left( n^j \bar{m}_j - \bar{n}_k m^k  \right) \Bigg]
\eea
Similar to the one-loop case, for the case of vanishing $B$ field, the terms in $m$ and $\bar{m}$ do not mix, and we can perform independent Poisson resummations to obtain 
\bea
\label{HigherGwithoutBfield}
Z_{cl.} &=& \sum_{n, \bar{n}, w, \bar{w}}
\textrm{exp} \Bigg[
-2\pi   \Bigg( G_{ij} (\bar{w}^i + \frac{1}{2}n^i ) \tau_2 (\bar{w}^j + \frac{1}{2}n^j)
+G^{ij} (w_i + \frac{1}{2}\bar{n}_i ) \tau_2  (w_j + \frac{1}{2} \bar{n}_j) \cr
&&+\frac{1}{4} (G_{ij} n^i \tau_2 n^j + G^{ij} \bar{n}_i \tau_2 \bar{n}_j ) \Bigg) 
+2\pi i \left(  \bar{n}_k \tau_1 \bar{w}^k - n^k \tau_1 w_k     \right) \Bigg]
\eea
This form is essentially the same as the one-loop case, and after factoring out the 
appropriate modes, we obtain the expected partition function. We now proceed to include the $B$ field. 
The modes in $m,\bar{m}$ are now mixed, but performing a Poisson resummation of $\bar{m}$, we 
obtain 
\bea
Z_{cl.} &=& \left( \textrm{Det}(G) \right)^{-g/2} \left( \textrm{Det}(\tau_2) \right)^{d/2} \sum_{n, \bar{n}, m, \bar{w}}
\textrm{exp} \Bigg[
-\frac{\pi}{2} G_{ik} m^i \frac{1}{\tau_2} m^k + 2\pi i \Bigg\{ B_{ik} \left(\bar{w}^k + D^k \right) - \frac{1}{2}\bar{n}_i \cr
&&+  \left[  (e^-_l e^-_i + e^+_l e^+_i ) n^l + (e^k e^-_i - e^k e^+_i ) \bar{n}_k \right] 
\tau_1 \frac{1}{4i\tau_2}
\Bigg\} m^i \cr
&&-2\pi G_{ij} (\bar{w}^i + D^i ) \tau_2 (\bar{w}^j + D^j) - \frac{\pi}{4} 
\Bigg[ (e^-_i n^i + e^j \bar{n}_j)
\left(  \tau_2 + \tau_1 \frac{1}{\tau_2} \tau_1 \right)(e^-_i n^i + e^j \bar{n}_j) \cr
&&+ (e^+_i n^i - e^j \bar{n}_j)\left(  \tau_2 + \tau_1 \frac{1}{\tau_2} \tau_1 \right)(e^+_i n^i - e^j \bar{n}_j)
\Bigg] \Bigg]
\eea
where we have defined 
$$
D^k = \left(  G^{rk} \bar{n}_r - B_{rm} G^{mk} n^r \right) \tau_1\frac{1}{2i\tau_2}  + \frac{1}{2}n^k,\qquad
e^{\pm}_i = e_i \pm B_{ij} e^j
$$
The term linear in $\bar{m}$ simplifies to read 
$$
K_i= B_{ik} (\bar{w}^k + \frac{1}{2}n^k ) - \frac{1}{2} \bar{n}_i + G_{li} n^l \tau_1\frac{1}{2i\tau_2} 
$$
After Poisson resumming $m\rightarrow w$, the instanton part of the partition function can be simplified to read
\bea
\label{HGeventualL}
Z_{cl.} &=&  \left( \textrm{Det}(\tau_2) \right)^{d} \sum_{n, \bar{n}, w, \bar{w}}
\textrm{exp} \Bigg[
-2\pi G^{ik} (w_i + K_i) \tau_2 (w_k + K_k) - 2\pi  G_{ij} (\bar{w}^i + D^i) \tau_2  (\bar{w}^j + D^j) \cr
&&-\frac{\pi}{4} \Bigg(  2(n_a  \left(
\tau_2 + \tau_1\frac{1}{\tau_2} \tau_1
\right)
   n^a + B_{am} G^{mk} B_{bk} n^a \left(
\tau_2 + \tau_1\frac{1}{\tau_2} \tau_1
\right)
 n^b ) + 2\bar{n}^a  \left(
\tau_2 + \tau_1\frac{1}{\tau_2} \tau_1
\right)
 \bar{n}_a \cr
&&- 4 B_{im}G^{mk} n^i \left(
\tau_2 + \tau_1\frac{1}{\tau_2} \tau_1
\right)
 \bar{n}_k \Bigg) \Bigg]
\eea
It is straightforward to verify that 
\eqref{HGeventualL} can be massaged into the form (again, just like in \eqref{Tfinal}, we suppress all indices to let the form of the expression be more revealing)
\bea
Z_{cl.} &=&  \left( \textrm{Det}(\tau_2) \right)^{d} \sum_{n, \bar{n}, w, \bar{w}}
\textrm{exp} \Bigg[
 \frac{i\pi}{2}  \left( (n+\bar{w} + B(n+\bar{w}) + w)\tau (n+\bar{w} + B(n+\bar{w}) + w) \right) \cr
&&+ \frac{i\pi}{2} \left( (\bar{n} - w -B\bar{w} + \bar{w}) \tau (\bar{n} - w -B\bar{w} + \bar{w}) \right) \cr
&&-\frac{i\pi}{2} \left( (\bar{n}-w-B\bar{w} - \bar{w})\bar{\tau}  (\bar{n}-w-B\bar{w} - \bar{w}) \right) \cr
&&-\frac{i\pi}{2}\left(  (-(n+\bar{w}) + B(n+\bar{w}) + w) \bar{\tau} (-(n+\bar{w}) + B(n+\bar{w}) + w) \right) \Bigg]
\eea
We can identify all left- and right-moving momenta as 
\bea
p^\alpha_{La} &=& e^m_a \left( w^\alpha_m + B_{mk} u^{\alpha k} + G_{mk}u^{\alpha k} \right) \cr
p^\alpha_{Ra} &=& e^m_a \left( v^\alpha_m - B_{mk} \bar{w}^{\alpha k} - G_{mk} \bar{w}^{\alpha k} \right) \cr
q^\alpha_{La} &=& e^m_a \left( v^\alpha_m - B_{mk} \bar{w}^{\alpha k} + G_{mk} \bar{w}^{\alpha k} \right) \cr
q^\alpha_{Ra} &=& e^m_a \left( w^\alpha_m + B_{mk} u^{\alpha k} - G_{mk}u^{\alpha k}   \right) 
\eea
where we have restored all the indices. Factoring out $p_R, q_L$, we obtain the correct partition function. Thus, we see from this section that the evaluation of the classical instanton contribution to the higher-genus partition function is rather straightforward when all the winding and momenta zero modes are defined with respect to the canonical basis of cycles of the Riemann surface. 

In the bulk of our discussion in all previous sections, we have been mostly deriving the classical instanton part of the doubled partition function. In the next section, we shall discuss the contribution of the string oscillators to the doubled partition functions. For the general toroidal background and even on higher-genus worldsheets, we shall see that the quantum part factorizes into a product of holomorphic and anti-holomorphic functions.  Imposing the constraint amounts to a neat removal of half of the holomorphic (and anti-holomorphic) functions, and thus we obtain the correct contribution to $Z$. We shall also see in Section \ref{SUSY} that for the supersymmetric theory, this is the case for the fermions as well. 

\section{On the contributions of the quantum oscillators to the partition function}
\label{Quantum}

In this section, we fill up the gap in all the previous sections where the focus was exclusively on understanding how half of the classical instantons or zero modes are retained consistently, so that upon chiral factorization of the Hamiltonian representation of the partition function, we recover the correct partition function. A definitive statement on the consistency of such an approach should of course be accompanied by checking that the quantum part of the partition function - due to the oscillator modes - can be reduced appropriately by chiral factorization. 

In the doubled formalism, we have emphasized that it is crucial that the Lagrangian can be written as a sum of doubled sets of kinetic terms of $P$ and $Q$. We have seen explicitly that this is true for the $\mathcal{N}=1$ worldsheet  and for the general toroidal background. Chiral factorization means that we keep the holomorphic part of $P$ and the anti-holomorphic part of $Q$. This can be done straightforwardly if the quantum part of the partition function contains $|F(\tau)|^2$ for some complex function $F(\tau)$. 

Nonetheless there is a rather important subtlety that arises when we think about higher-genus string worldsheets that is not so apparent from a one-loop computation. For each classical instanton sector, recall that after a Poisson resummation, we obtain some prefactors containing the determinants of the target space metric tensor and the imaginary part of the period matrix. Explicitly,
\be
\label{Zclass}
Z_{cl.} = \left(  \textrm{Det} (\tau_2) \right)^{\frac{d}{2}} \left(  \textrm{Det} (G) \right)^{-\frac{g}{2}} \sum_{p_L,p_R} \textrm{exp} \left[\frac{i\pi}{2} ( p_L)_\alpha \tau_{\alpha \beta} (p_L)_\beta 
-\frac{i\pi}{2} ( p_R)_\alpha \bar{\tau}_{\alpha \beta} (p_R)_\beta\right] 
\ee
On the other hand, the quantum part of the partition function that takes into account the oscillator modes reads
\be
\label{Zquant}
Z_{q.} = \frac{\sqrt{\textrm{Det}(G)}}{(\textrm{Det} (\tau_2))^{\frac{d}{2}} |\eta_g|^2 }
\ee
where $\eta_g$ is an appropriate generalization of the Dedekind eta-function that comes from computing the determinant of the scalar Laplacian on higher-genus surfaces. We note that the factor of $\sqrt{\textrm{Det} (G)}$ originates from the zero mode of $X^\mu$ and measures the volume of the target space. Taking the product of $Z_{cl.}$ and $Z_{q.}$, we see that while $\textrm{Det}(\tau_2)$ cancels away, there is a remnant factor of $\left( \textrm{Det}(G) \right)^{(1-g)/2}$. This factor implies that on its own, the partition function is not T-duality invariant. As is well-known, this factor of $\textrm{Det}(G)$ precisely matches the T-duality covariance of the dilaton. Recall that the complete closed string partition function is a sum over all genus. Schematically, we can write 
\be
Z_{string} = \sum_{g=0}^\infty e^{(1-g)\Phi} Z_g
\ee
where $Z_g$ is $Z_{cl.} Z_{q.}$ at each genus $g$. Under T-duality, the dilaton transforms such that the following quantity is a T-duality invariant.
\be
\label{dilatonTrule}
\Phi + \textrm{Log} \sqrt{ \textrm{Det} \,G} 
\ee
Thus, we see that at each genus, T-duality is a symmetry. In the doubled formalism, we have seen that in computing $Z_{cl.}$, the factor of $\textrm{Det}(G)$ cancels away and is absent after Poisson resumming the winding modes along the Euclidean time direction. (Note that we still have the factor of $\left(\textrm{Det}(\tau_2)\right)^d$. ) On the other hand, for the quantum part, since the metric $G$ is just $\frac{1}{2}\times \mathds{1}_{2d\times 2d}$, we retrieve a factor of some power of $2$ which cancels away an identical factor in $Z_{cl}.$ This leaves us simply with the function $|\eta_g|^2$ in the denominator and the sum of all left and right-moving momenta. We thus observe that there is no dependence on $\textrm{Det}(G)$. 
This point suggests that in the doubled formalism, the scalar field that plays the role of the dilaton should be invariant under T-duality. 

Indeed, as was explained in \cite{Hull:2006va}, this is the case. In the doubled theory, the parent T-duality invariant dilaton field $\phi$ presents itself in a Fradkin-Tseytlin term that reads 
\be
\label{FradDilaton}
S_{FT} = \int d^2 \sigma \sqrt{h} \phi R.
\ee
where $R$ is the worldsheet's Ricci scalar. The relation to the usual dilaton field that transforms under \eqref{dilatonTrule} is manifest after integrating over the auxiliary fields as contained in the term $\frac{1}{4} G^{ij} D_i \wedge \ast D_j$. This yields a determinant which replaces the $\phi$-term in \eqref{FradDilaton} with \cite{Tseytlin:1991wr,Schwarz}
\be
\label{DoubledDilaton}
\Phi = \phi - \frac{1}{2} \textrm{log}(\textrm{Det}(G_{ij})). 
\ee
T-duality transformation rule allows us to identify $\Phi$ as the dilaton field. To relate this fact to
our calculations, we note that integrating out the auxiliary fields in the Lagrangian picture, the spectrum of the theory is what we should obtain after performing the chiral factorization. In the latter procedure, within the framework that we considered, there is no principle that dictates us to put in a factor of $\left(\textrm{Det}(G)\right)^{-g/2}$ in $Z_{q.}$ after imposing the constraint. Although the final result is consistent with a T-duality invariant $\phi$, it would be important to develop a thorough understanding of how the chiral factorization procedure applies to $\phi$. It was mentioned in \cite{Hull:2006va} that the expectation value of $e^{-\phi}$ is the string field theory coupling constant of \cite{Hull:2006va}. It appears that a more complete and natural language for the higher-genus worldsheet theory lies in string fields \cite{Kugo,Alvarez2}. Our results only demonstrates consistency for the theory as a perturbation series in the expectation value of $\phi$. 

Finally, before we proceed to the next section, let us briefly describe $|\eta_g|$ that we alluded to in \eqref{Zquant}. We simply wish to point out that it is of the form $|F(\tau)|$ as this is necessary for chiral factorization to work. We leave pedagogical details to Appendix A while briefly pointing a couple of crucial facts for our argument. Let us first write the one-loop result in a slightly different form. For a flat torus with metric
$ds^2 = \frac{|dz|^2}{\tau_2}$, $\nabla^2 (\tau) = -\tau_2 \frac{\partial^2}{\partial z \partial \bar{z}}$. 
After excluding the zero eigenvalues, and using zeta function regularization, we can write 
\be
\label{torusdet}
\textrm{Det}'\left( \nabla^2 \right) = 4\tau_2 |\eta(\tau)|^4 = 4 \tau_2
e^{-\frac{S(\tau)}{12\pi}} \prod_{m=1}^{\infty} \left| 1-q^m \right|^4
\ee
where 
\be
\label{Ltorus}
S_L(\tau) = \frac{i}{2} \int \int_D \left|    \frac{\partial \textrm{log} \rho}{\partial w}   \right|^2 dw    d\bar{w} = 4\pi^2 \tau_2,\,\,\, \rho(w) = \frac{1}{4\pi^2 \tau_2 |w|^2},\,\,\,
D= \left\{ w \in \mathbb{C}: |q| < |w| \leq 1 \right\}
\ee
This is motivated by describing the torus as the Schottky uniformization of the elliptic curve $\Gamma \backslash \mathbb{C}$ where $\Gamma$ is the cyclic group generated by the dilation $w \rightarrow qw$, with $D$ as the fundamental region, and $w=e^{2\pi i z}$, with the metric $ds^2 = \rho (w,\bar{w}) |dw|^2$. The function $S_L(\tau)$ is known as the classical Liouville action defined on the Schottky space that we just described. 

In the case of higher-genera worldsheets, it is useful to write the Laplacian determinant in a form that manifestly admits chiral factorization. As a common approach \cite{Phong}, the determinant is computed for a Riemann surface $X$ by regarding it as the upper half plane $\mathds{H}$ uniformized by some Fuchsian group, in which case it depends on the Selberg zeta function which is expressed in terms of real quantities and hence it may not be obvious as to how we can holomorphically factorize the determinant if we wish to show the equivalence between the doubled and ordinary sigma model. But we point out that a result due to Zograf \cite{Zograf} yields an expression for the determinant equivalent to the chiral factorization of the Selberg zeta function. This relies on a description of $X$ by the quotient of a suitable domain of the Riemann sphere by Schotty groups. As explained in the Appendix, it is useful to think of the commutative diagram
\begin{center}
\begin{tikzpicture}[description/.style={fill=white,inner sep=2pt}] 
\matrix (m) [matrix of math nodes, row sep=3em, column sep=2.5em, text height=1.5ex, text depth=0.25ex] { \mathds{H} & & \Omega \\
& X & \\ };
\path[->,font=\scriptsize] 
(m-1-1) edge node [above] {$ J $} (m-1-3)
edge node [below left] {$ \pi_f $} (m-2-2) 
(m-1-3) edge node [below right] {$ \pi_s $} (m-2-2);
\end{tikzpicture}
\end{center}
where $\pi_f : \mathds{H} \rightarrow X$ and $\pi_s : \Omega \rightarrow X$ are the quotient maps corresponding to the Fuchsian and Schottky groups respectively, and $J$ is a covering map (see Appendix).

We can now write down the higher-genus generalization straightforwardly. Following \cite{Zograf}, 
the determinant of the scalar Laplacian on the higher genus surface can be written as 
\be
\label{SchottkyDet}
\textrm{Det}' \nabla^2 = \left( \textrm{Det} \tau_2 \right) \textrm{exp} \left( -\frac{S_L}{12\pi} \right) |F|^2,\qquad 
F = \prod_{\{ \gamma \}} \prod_{m=0}^\infty \left(  1 - q^{1+m}_\gamma \right)
\ee
where $q_\gamma$ is the multiplier of $\gamma \in \Gamma$ and $\gamma$ runs over all distinct primitive\footnote{Primitive elements refer to those which cannot be obtained by some powers of other elements of the group.} conjugacy classes in $\Gamma$ excluding the identity, and $S_L$ is the classical Lioville action defined on the higher-genus Schottky space (for the torus, $S_L = -\pi \textrm{Log} |q|^2$). In the context of chiral factorization, imposing the constraint is effectively equivalent to taking the square root of the determinant, retaining the factors $e^{-\frac{S_L}{24\pi}} F(\tau)$ and $e^{-\frac{S_L}{24\pi}} \overline{F(\tau)}$ for the oscillator contributions of the holomorphic $P$ and anti-holomorphic $Q$  respectively. We refer the reader to Appendix A and references contained in it for a more pedagogical explanation.

\section{$\mathcal{N}=1$ Worldsheet Supersymmetry}
\label{SUSY}

Let us attempt to generalize our results to the doubled formalism with supersymmetry. We first briefly review some basic aspects of the supersymmetric doubled sigma model following \cite{Hull:2006va}. The $\mathcal{N}=1$ generalization of the ordinary sigma model in $(1,1)$ superspace reads \cite{Gates}
\be
\label{susystringaction}
S = \int d^2 \sigma d^2 \theta \left(  G_{\mu\nu} C^{rs} + B_{\mu \nu} \gamma^{rs} \right)
D_r \Phi^\mu D_s \Phi^\nu
\ee
where $\Phi^\mu (\sigma, \theta)$ is a superfield on the superspace worldsheet with coordinates
$\sigma^\alpha, \theta^r$, the latter being real anti-commuting coordinates that transform as a worldsheet spinor, $r=1,2$ and $D_r$ are the supercovariant derivatives. We work with Majorana spinors, with $\theta_{\alpha}, \alpha=1,2$ parametrizing the $\mathcal{N}=1$ worldsheet supersymmetry. We choose our Dirac matrices to be 
\be
\label{Dirac}
\rho^0 = \sigma_2, \rho^1 = i\sigma_1, \rho^2 = \sigma_3
\ee
where $\sigma_i$ are the Pauli matrices and $\bar{\theta}_\alpha = \theta_\beta \rho^0_{\beta \alpha}$, and our superfields and their covariant derivatives read 
\bea
\label{superfields}
&&\Phi^I = \mathbb{X}^I + \bar{\theta} \psi^I + \frac{1}{2} \bar{\theta} \theta f^I, \cr
&& D_\alpha \Phi^I = \psi^I_\alpha + \theta_\alpha f^I - i \left(  \rho^m \theta \right)_\alpha \partial_m \mathbb{X}^I + \frac{i}{2}\partial_m \left( \rho^m \psi^I \right)_\alpha \bar{\theta} \theta.
\eea
Also, $C$ is the charge conjugation matrix, and $\rho_2$ the chirality operator satisfying
$$
C^{rs} = \epsilon^{rs}, \qquad \gamma = C \rho_2.
$$
Correspondingly, the action of the supersymmetric doubled sigma model reads 
\bea
\label{susydoubledaction}
S &=& \frac{1}{2} \int d^2 \sigma d^2 \theta \,\,\, \frac{1}{4} H_{IJ} C^{rs} \left[ D_r \mathbb{X}^I + \mathcal{A}^I_m D_r Y^m \right]\left[ D_s \mathbb{X}^J + \mathcal{A}^J_n D_s Y^n \right] \cr
&&-\frac{1}{2} \gamma^{rs} L_{IJ} \left[ D_r \mathbb{X}^I + \mathcal{A}^I_m D_r Y^m \right] \mathcal{A}^J_s + \mathcal{L} (Y).
\eea
As in the bosonic case, we need to augment it with an appropriately supersymmetrized topological term which reads 
\be
\label{susytop}
\mathcal{L}_{top.} = \frac{1}{2} \gamma^{rs} \Omega_{IJ} D_r \mathbb{X}^I D_s \mathbb{X}^J
\ee
whereas the constraint is supersymmetrized to be
\be
\label{constraint}
\left[ D_s \mathbb{X}^J + \mathcal{A}^J_n D_s Y^n \right] = S \left(\gamma_3 \right)_{sr} \left[ D_r \mathbb{X}^J + \mathcal{A}^J_n D_r Y^n \right]
\ee
Below we shall restrict ourselves to the case where the source terms $\mathcal{A}$ vanish. After expanding each superfield in its constituents and keeping only terms in $\bar{\theta} \theta$ since $\int d^2 \theta (\bar{\theta} \theta) = 1$, we find 
\bea
\label{susydoubledaction}
\mathcal{L} &=& H_{IJ} \left( \partial^a \mathbb{X}^I \partial_a \mathbb{X}^J + i \bar{\psi}^I \rho^m \partial_m \psi^J + f^I f^J \right) \cr
&&\qquad \qquad + i\pi \Omega_{IJ} \left[ \partial_t \mathbb{X}^I \partial_\sigma \mathbb{X}^J + i\psi^I_2 (\partial_t - \partial_\sigma) \psi^J_2 + i\psi^J_1 (\partial_t + \partial_\sigma) \psi^I_1          \right]
\eea
We note that since $\Omega$ is anti-symmetric, the fermionic bilinear terms are total derivatives. They do not contribute to the classical instanton sector of the partition function. 
On-shell, the auxiliary fields $f$ vanish (in general, this is not true when we consider fibrations with generic $O(d,d;\mathbb{Z})$ transition functions), and the real fermions satisfy the equations of motion 
\be
\label{fermionEOM}
(\partial_t - \partial_\sigma) \psi_2 = 0,\,\,\, (\partial_t + \partial_\sigma) \psi_1 = 0
\ee
We observe from \eqref{susydoubledaction} that the topological term contains fermionic bilinears which are total derivatives. We should of course check the constraint terms too to ensure that the bosonic contraints are recovered. Indeed they are, and we also have similar constraints for the fermions which read\footnote{There is also another constraint that arises from the quadratic $\theta$ term which we find to be equivalent to the fermion's equation of motion in \eqref{fermionEOM}.}
\be
\label{fermionconstraint}
\psi^I_1 = -S^I_J \psi^J_1, \qquad \psi^I_2 = S^I_J \psi^J_2
\ee
Recall that when we compute the partition function in earlier sections, the fact that we are able to express the non-topological part of the Lagrangian as a sum of two kinetic terms (for $P$ and $Q$) is crucial in arguing that the quantum fluctuations that arise from the oscillators admit a chiral factorization, and that we can effectively take the square root of this portion of the doubled partition function. With the fermions, we can do likewise. For each of $\psi_{1,2}$, we find that if we define
\bea
\varphi^a &=& \left( e^a_i + B_{ij} e^{ja} \right) \psi^i + e^{ja} \tilde{\psi}_j \cr
\eta^a &=& \left( e^a_i - B_{ij} e^{ja} \right) \psi^i - e^{ja} \tilde{\psi}_j,
\eea
then in these variables, the constraints translates into simply $\varphi^a_1 = \eta^a_2 = 0$. Note that just like their bosonic counterparts, these fields are valued in the tangent space indices. Formally, this projection into half the degrees of freedom is an endomorphism of the target tangent bundle. The fermionic part of the Lagrangian can now be simply expressed as
\be
\mathcal{L}_{fermions} = \frac{1}{2} \varphi^a_2 \partial_z \varphi^a_2 + \frac{1}{2} \varphi^a_1 \partial_{\bar{z}} \varphi^a_1 + \frac{1}{2} \eta^a_2 \partial_z \eta^a_2 + \frac{1}{2} \eta^a_1 \partial_{\bar{z}} \eta^a_1
\ee
Like the bosonic fields, the fact that within the Lagrangian, the fermionic fields can be expressed as a sum of kinetic terms is crucial for establishing the equivalence of the doubled sigma model with the ordinary one. Even though the fermionic terms do not contribute to the classical instanton part, the quantum part can be written after imposing either periodic or anti-periodic boundary conditions along each homology cycle for each chiral fermion. These determinants for the chiral fermions including their higher-genera counterparts are discussed in for example the seminal papers \cite{Alvarez, Verlinde, Phong}. 
For the fermions, imposing the constraint after evaluating the full partition function is even a simpler process. Since $\varphi^a_1, \eta^a_2$ are set to vanish, what is left is half of the doubled contribution.

\section{Conclusion}

In this work, we have demonstrated that the ordinary closed string partition function in a general toroidal compactification, including their shift orbifolds, can be reproduced by interpreting
the constraint equation in Hull's doubled geometry formalism as that of chiral scalars. This furnishes a non-trivial consistency check for the quantum equivalence to the ordinary theory.\footnote{Please see \cite{Berman:2007} for another crucial quantum check of the doubled sigma model.}  Moreover, we presented explicit computations showing that this equivalence holds tight at all higher worldsheet genus (in which Schottky uniformation of Riemann surfaces appears to play a nice role), and also for the case of $\mathcal{N}=1$ worldsheet supersymmetry. 

The partition function at a fixed toroidal metric can be decomposed into a part that sums up the string's zero modes and a part that takes into account the quantum fluctuations of the oscillators. For the instanton part, the topological term in the doubled sigma model plays a crucial role in ensuring that chiral factorization can allow us to consistently reduce the doubled partition function to the corresponding instanton sums in the conventional string theories, after a relabelling of winding numbers. This term contributes certain relative signs in the instanton sum, and as explained in \cite{Hull:2006va}, it is needed to completely gauge away the auxiliary degrees of freedom. For the quantum part, the consistency of the chiral factorization procedure relies on a couple of prerequisites. Firstly the Lagrangian can be written in a form manifestly independent of the target space metric in terms of the new chiral variables, and secondly that it contains the sum of these variables' kinetic terms. We illustrated clearly that this can be done for the general toroidal background and for fermions when the worldsheet is endowed with $\mathcal{N}=1$ supersymmetry. 

We have also shown that the interpretation of the constraint equations as that of chiral scalars works for string worldsheets of higher genus. Comparing the instanton sum to the ordinary case, we find that it differs from the latter by being independent of the determinant of the target space metric. This in turn implies that it is only consistent with an accompanying dilaton field $\phi$ that is T-duality invariant. Indeed, this is the case as was explained in \cite{Hull:2006va} and \cite{Berman}. However, our technique does not empower us to understand how to impose the constraint on the dilaton $\phi$. In \cite{Hull:2006va}, it was shown that when the auxiliary degrees of freedom are integrated away, the dilaton $\phi$ can be related to the ordinary dilaton field with the correct T-duality transformation rules. We do not know how or whether the principle of chiral factorization can allow us to derive this relationship. At this level of our approach, we can at best make a statement on the consistency of the doubled formalism as a perturbation series in $\phi$. At genus one, this problem is not as apparent since the instanton sum is manifestly T-duality invariant. 

Indeed, as much as we have demonstrated the consistency of interpreting the constraints as that arising from chiral scalars, we have also learnt to be aware of the limitations of such an approach. Apart from the point about $\phi$ for the higher-genus worldsheets, we also found that the doubled formalism does not seem to `geometrize' asymmetric orbifold action. In our exploration of $\mathbb{Z}_N$ shift orbifolds, chiral factorization reproduces the expected partition function when the shift is geometric and thus acts symmetrically on the left and right-movers. For asymmetric shifts, we find that we need additional pairing phase factors identical to the ordinary case. These factors do not have a clear origin in the Lagrangian picture, but after Poisson resumming to be in the Hamiltonian picture, they arise naturally when we compute the insertion of the translation operator. Although we have demonstrated that we still recover the expected result, it would be nice if these phase factors arise from a geometrical origin in the doubled sigma model. One of the main purpose of the doubled formalism is to furnish a framework for T-folds. This generically involves asymmetric orbifolds when the T-duality is a symmetry of the background's moduli space. It is known that modular invariance imposes conditions on the allowed asymmetric orbifold action, and more generally for T-folds, the legitimate $O(d,d;\mathbb{Z})$ valued transition functions of the torus fibration. The simplest T-fold which involves an asymmetric $\mathbb{Z}_2$ reflection on $S^1$ target space was previously investigated in \cite{HackettJones}, where it was found that to construct a modular-invariant partition function, we need to impose the values of the eigenvalues of the T-duality operator  (in the higher-dimensional case, this corresponds to the $O(d,d;\mathbb{Z})$ element that inverts all the circles' radii). These conditions were taken from the beautiful work of \cite{Simeon} but we think that it would be very interesting to see them arising naturally from the doubled formalism. Our result for the $\mathbb{Z}_N$ shift orbifolds suggest that a supplementary analysis in the operator formalism as, for example, undertaken in \cite{HackettJones} is necessary to develop a deeper understanding of modular invariance when asymmetric orbifold action is involved. 

Throughout our work, we have appeared to be able to demonstrate the principle of chiral factorization rather smoothly to the general toroidal background and on higher-genus worldsheets. That this procedure falls into place so delicately is related to the underlying fact that in the doubled sigma model, $X$ and $\tilde{X}$ represent the Fourier transform duals of the momentum and winding numbers. A natural future work would be to consider T-folds with non-trivial T-duality transition functions. Will the principle of chiral block decomposition continue to play some role in these non-geometric backgrounds? In general, quantization of chiral scalars contains many subtleties (see \cite{Witten:1996hc} for an elucidating discussion) and a more thorough analysis is warranted. We hope to report our progress in these directions in the near future \cite{future}.

\section*{Acknowledgments}

It is a pleasure to thank David Berman for correspondence on his work in \cite{Berman} and for reminding me about Witten's work in \cite{Witten:1996hc}. I am grateful to Jan de Boer, Ori Ganor, and Masaki Shigemori for inspirations on various topics related to duality twists and non-geometric backgrounds. I acknowledge support from the Foundation for Fundamental Research on Matter (FOM) which is part of the Netherlands Organization for Scientific Research.

\appendix
\label{AppA}
\section{Chiral factorization of Laplacian Determinant from Schottky Uniformization of Riemann surfaces}

Compact Riemann surfaces admit several different descriptions, in terms of elliptic curves and uniformizations by Fuchsian and Schottky groups. Below, we briefly outline some essential points on the uniformization of a Riemann surface $X$ of genus $g \geq 1$ by Schottky groups. Let $\alpha_1, \ldots \alpha_g, \beta_1,\ldots, \beta_g$ be the generators of $\pi_1 (X)$. They form the canonical homology basis of $X$. Let $\mathcal{N}$ be the smallest normal subgroup in $\pi_1(X)$ that contains $\alpha_1, \ldots, \alpha_g$. 
By the classical retrosection theorem of Koebe\footnote{See for example \cite{gray} and references therein.}, there exists a Schottky group, 
$$
\Gamma \sim \pi_1(X) / \mathcal{N}
$$
with a domain  $\Omega$ of $\mathbb{C} \cup \infty$ such that $X \sim \Omega / \Gamma$. This group is marked by generators
$L_1, L_2, \ldots L_g$ corresponding to the cosets $\beta_1 \mathcal{N}, \beta_2 \mathcal{N}, \ldots, \beta_g \mathcal{N}$. The generators $L_i$ are complex projective transformations which can be visualized
as follows. We begin with the canonical cycles $\{ \alpha_i, \beta_i \}$ of the Riemann surface. If we break this diagram along the $\alpha$ cycles, and then map it conformally onto the region of the complex plane exterior to $g$ pairs of circles, then for each pair of circles $(\alpha_i, \alpha'_i)$, we can associate a complex projective transformation $L_i$ that maps the circles' circumferences onto one another and identifying the points on them. The $\alpha$ cycles surround the circles while the $\beta$ cycles take us between the circles. The $L_i$'s and their inverses can be multiplied by one another in all combinations (apart from the restriction that $L_i$ be not placed next to its inverse), and thus we obtain an infinite group of projective transformations as a representation of $\pi_1(X)$. The Schottky group $\Sigma_g$ of genus $g$ is freely generated by the $L_i$ which can always be conjugated to be of the form 
\be
\label{Schottkyeqn1}
\frac{L_i(w) - a_i}{L_i(w) - b_i} = q_i \frac{w-a_i}{w-b_i}
\ee
where $a_i, b_i$ are the attractive and repelling fixed points of the transformations, $q_i$ is  the multiplier satisfying $|q_i| <1$. In a $2\times 2$ matrix representation, $\Gamma$ acts on $\Omega$ via matrices $\gamma$ of which absolute trace is bigger than 2 (these are the `loxodromic' elements). For example for the generators $L_i$,
\bea
&& \gamma_i =  \left( \begin{array}{cc} A & B \\ C & D \end{array} \right) = 
\left(  \begin{array}{cc} a_i & b_i \\ 1 & 1 \end{array} \right) \left(  \begin{array}{cc} 1/\sqrt{q_i} & 0 \\ 0 & \sqrt{q_i} \end{array} \right) \left(  \begin{array}{cc} a_i & b_i \\ 1 & 1 \end{array} \right)^{-1} 
,\cr
&&L_i (w) \equiv \frac{Aw + B}{Cw +D}, \,\,\, w \in \Omega.
\eea
The generic element $V$ is then some product of the $L_i$'s. To summarize, the Riemann surface of genus $g$ can be constructed as taking $2g$ disjoint circles $\{ \alpha_i, \alpha'_i
\}$ chosen such that they are identified under $L_i (\alpha_i) = \alpha'_i$. This obviously creates $g$ handles on the Riemann sphere, and thus under the quotient of the Schottky group, we have a genus-$g$ Riemann surface. We end up with the following qualitative picture. The genus-$g$ Riemann surface is an $S^2$ with $g$ holes cut out and identified pairwise. The fixed points $\{ a_i, b_i \}, i =1,\ldots g$ in 
\eqref{Schottkyeqn1} represent handle positions, while the multipliers $q_i$ represent the length and the twist of the handle.  

After identification, the Riemann surface is then the space exterior to all the circles, each of which contain the fixed points. For example, for genus one, the torus can be described by the set of points $\{ w \in \mathbb{C} \cup \infty \vert w \sim qw \}$, for some $q=e^{2\pi i \tau}$. Setting $|q|<1$ amounts to choosing a positive $\textrm{Im}(\tau)$. The fundamental region can be chosen to be the annulus $1 \geq |w| > |q|$ with the boundaries identified to give a torus topology. The inner boundary circle has a fixed point at the origin, while the outer boundary circle has a fixed point at infinity. This admits a clear generalization for the surfaces of higher genera. 

Indeed, by conjugation in the group of all projective transformations, we can fix $a_1 = 0, b_1 = \infty, a_2 = 1$. Hence, for the Schottky group $\sigma_g$ of genus $g\geq 2$, the map
$$
\sigma_g \rightarrow \left( a_3, \ldots , a_g, b_2, \ldots b_g, q_1, \ldots, q_g \right) \in
\mathbb{C}^{3g-3}
$$
yields an isomorphism between the set of Schottky groups and some subset of $\mathbb{C}^{3g-3}$ which is known as the Schottky space of genus $g$ of complex dimensions $3g-3$. \footnote{Although this is equivalent to the dimension of the moduli space of Riemann surfaces, it remains for one to specify the correct Weil-Petersson measure in the moduli space of the string path integral. See for example \cite{DiVecchia} for the Weil-Petersson measure ($dV_{W.P.}$) in terms of Schottky parameters.} Schottky spaces are not known explicitly in general, but in principle the $3g-3$ moduli parameters are to be integrated over the fundamental domain of the appropriate modular group. The domain lies within the Schottky space. Although it is difficult to determine it explicitly (just as in the Schottky problem), the genus-one case is easily describable. For $T^2$, the only free complex parameter is the multiplier $q=e^{2\pi i \tau}$ where $\tau$ is the complex structure. The fundamental domain on the upper half plane translates into a bounded region in the $q$-plane, with the degenerate point $\tau = i\infty$ being mapped onto the origin as indicated in Figure 1.

\begin{figure}[h]
\centering
\mbox{\subfigure[$\tau$-plane]{\includegraphics[width=0.45\textwidth]{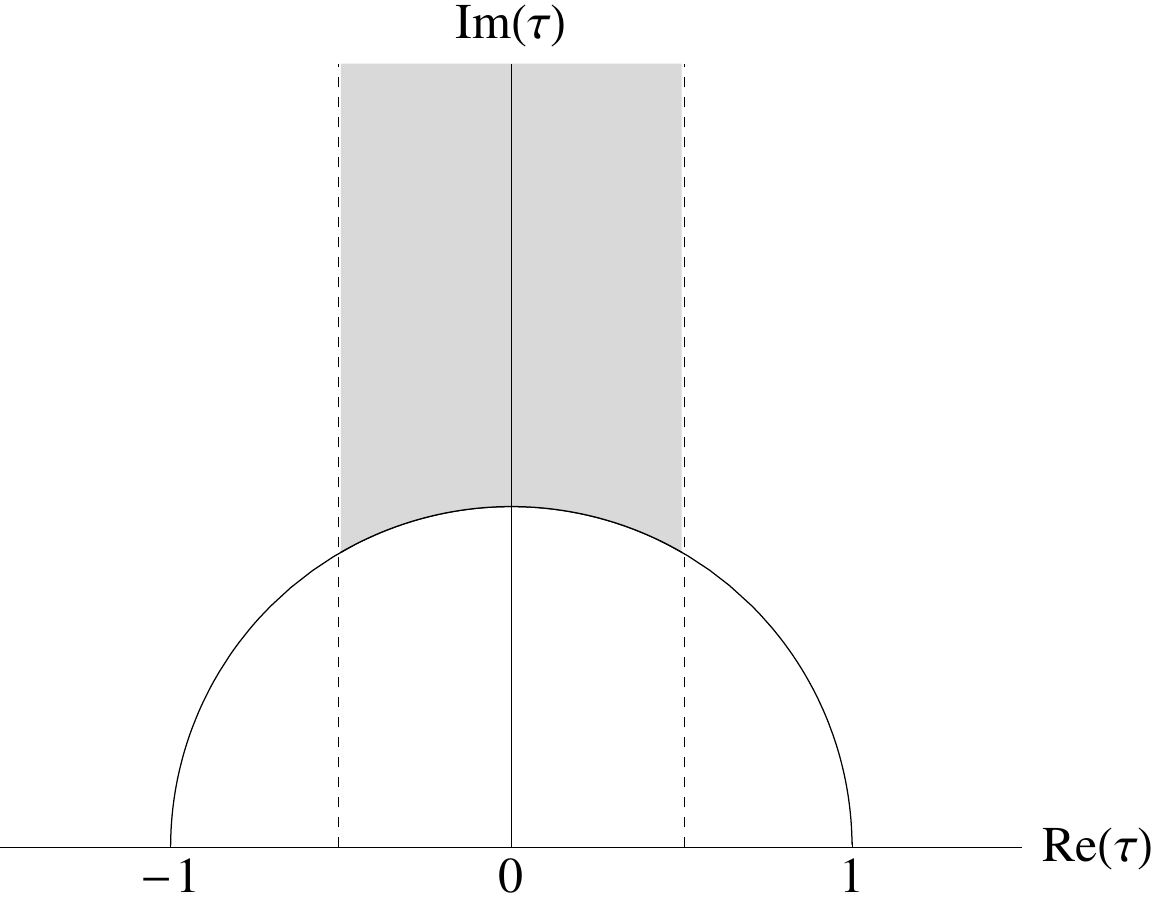}}\quad \quad \quad \quad \quad
\subfigure[$q$-plane]{\includegraphics[width=0.40\textwidth]{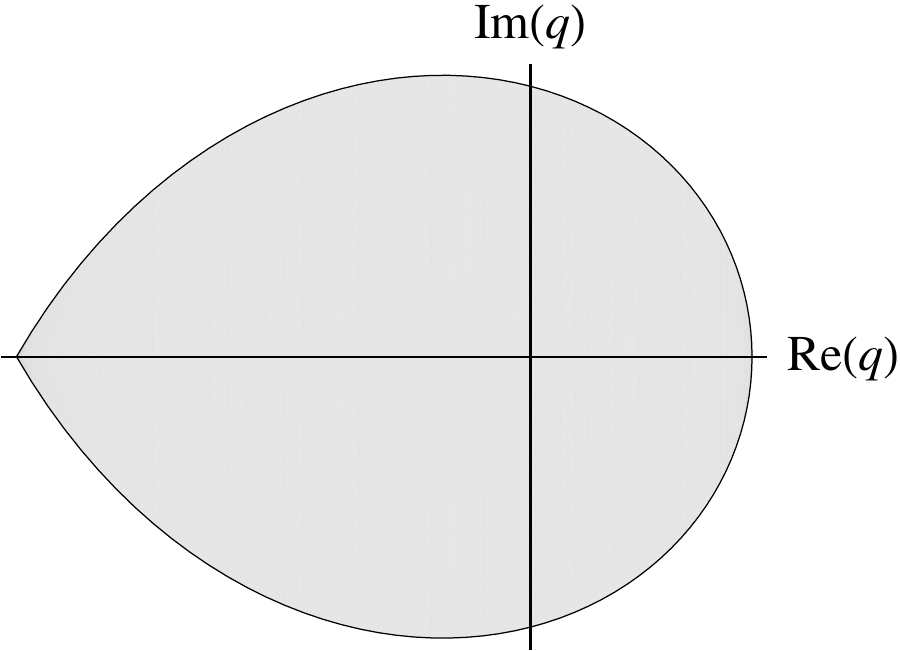}}}
\caption{In this figure, we display a choice of the fundamental domain of the moduli space for the case of genus one. Under the map $q=e^{2\pi i \tau}$, the shaded fundamental region translates into a bounded domain in the Schottky space (the $q$-plane). The vertices at $\tau = e^{\frac{\pi i}{3}} \sim e^{\frac{2\pi i}{3}}$ of Fig.1(a) map onto the crunode in Fig. 1(b), while the third at $\tau = i\infty$ maps onto the origin.   } \label{Fig1}
\end{figure}

Since the Laplacian determinant contains the period matrix of the Riemann surface, we should relate it explicitly to Schottky parameters too. Fortunately, this was explained beautifully by Mandelstam in \cite{Mandelstam} of which we quickly review below (see also \cite{Sakai}). Let us start with the abelian differentials $\omega_s \equiv d\nu_s$, the holomorphic one-forms that span $H_{(1,0)} \left( M_g, \mathds{Z} \right)$ that are stated earlier in \eqref{periodmatrix}. In \cite{Mandelstam}, Mandelstam found the following expression of $\nu_s$ in terms of Schottky parameters as follows\footnote{See also \cite{Cremmer2} for an earlier derivation and for a nice discussion for the modular invariance of the Weil-Petersson measure in Schottky parameters.}
\be
\nu_s (w) = \frac{1}{2\pi i} {\sum_i}^{(s)} \textrm{Log} \left(  \frac{w - V_i a_s }{w - V_i b_s}   \right) 
\ee
where ${\sum_i}^{(s)}$ denotes a summation over all values of $i$ except for those for which $V_i$ has a factor $L_s$ or its inverse at its RHS. So we see that each abelian differential can be nicely described as an infinite sum over the Schottky group. By examining transformation properties of $\nu_s$, we can deduce the period matrix to read \cite{Mandelstam, Cremmer2, Sakai} 
\bea
\tau_{\alpha \beta} &=& \frac{1}{2\pi i} \left(  {\sum_i}^{(\alpha ,\beta)} \textrm{Log} \frac{(a_\beta - V_i a_\alpha)(b_\beta - V_i b_\alpha)}{(a_\beta - V_i b_\alpha)(b_\beta - V_i a_\alpha)}  \right),\,\,\, \alpha \neq \beta
\qquad  \cr
\tau_{\alpha \alpha} &=& \frac{1}{2\pi i} \left( \textrm{Log}(q_\alpha) + 
{\sum_{i\neq \mathds{I}}}^{(\alpha ,\alpha)} \textrm{Log} \frac{(a_\alpha - V_i a_\alpha)(b_\alpha - V_i b_\alpha)}{(a_\alpha - V_i b_\alpha)(b_\alpha - V_i a_\alpha)}  \right).
\eea
where ${\sum}^{(\alpha,\beta)}$ is the summation over all elements of the Schotty group
excluding values of $\alpha, \beta$ such that $V_i$
has $L_\alpha$ (or its inverse) as its right-most member or $L_\beta$ (or its inverse) as its left-most member. 

Having reviewed how Schottky uniformation works in principle, we can now proceed to discuss the relevance of this description to the Laplacian determinant. A more common approach to write down the determinant is related to Fuchsian uniformization of the Riemann surface (see for example \cite{Phong}). A natural well-known function that arises in this context is the Selberg zeta function $Z(s)$ which reads 
$$
Z(s) = \prod_{\{\gamma\}} \prod^{\infty}_{m=0} \left( 1 - \lambda^{s+m}_\gamma \right)
$$
where $\{ \gamma \}$ runs over all distinct nontrivial primitive conjugacy classes in a Fuchsian group $\Gamma_f$ uniformizing the Riemann surface now described as $\mathds{H}/\Gamma_f$. Just as in the case of Schottky uniformization, our restriction to smooth and compact surfaces translates to the restriction of $\Gamma_f$ to hyperbolic elements. Each $\lambda_\gamma$ plays a role similar to that of a multiplier, and admits a geometrical meaning as closed geodesics of $\mathds{H}/ \Gamma_f$ corresponding to conjugacy classes
of hyperbolic elements. Explicitly, $\lambda_\gamma = e^{2l_\gamma}$ where $l_\gamma$ is the length of $\gamma$ in the hyperbolic metric. Although it is defined initially for $\textrm{Re}(s) >1$, $Z(s)$ admits an analytic continuation to the entire $s$-plane and the Laplacian determinant reads
\be
\label{SelbergZ}
\textrm{Det} (\nabla^2) = e^{c_0 (2g-2)} Z'(1)
\ee
for some constant $c_0$. As pointed out in for exampe \cite{Phong}, when we write down the determinant as \eqref{SelbergZ}, it is difficult to perform some form of chiral splitting, since the hyperbolic geometry and the Selberg zeta functions are defined by real quantities. 

In demonstrating the equivalence of the doubled sigma model to the ordinary one, we have interpreted Hull's constraints as that of chiral scalars. As we have observed in the main text, various quantities that the partition function comprises of should admit chiral splitting, and thus it is crucial that the determinant be holomorphically factorizable. Remarkably, it turns out that there is a solution to this problem by considering the Schottky uniformization of the Riemann surface. It was proven by Zograf in \cite{Zograf} when expressed in Schottky parameters, the determinant can be chirally factorized and written in the form of equation \eqref{SchottkyDet} which we repeat below for reading convenience.
\be
\textrm{Det}' \nabla^2 = \left( \textrm{Det} \tau_2 \right) \textrm{exp} \left( -\frac{S_L}{12\pi} \right) |F|^2,\qquad 
F = \prod_{\{ \gamma \}} \prod_{m=0}^\infty \left(  1 - q^{1+m}_\gamma \right)
\ee
where $q_\gamma$ is the multiplier of $\gamma \in \Gamma$, and $\{ \gamma \}$ runs over all distinct primitive conjugacy classes in $\Gamma$ excluding the identity and $S_L$ is the classical Liouville action that measures the holomorphic anomaly. This action is the K$\ddot{\textrm{a}}$hler potential of the Weil-Petersson metric on the Schottky space \cite{Teo} and be written as \cite{McIntyre}
\bea
\label{Liouvilleaction}
S_L &=& \iint_{F_d} \frac{i}{2} dw \wedge d\bar{w} \left(   \left| \frac{\partial \textrm{Log}(\rho)}{\partial w} \right|^2 + \rho \right) + 4\pi \sum_{k=2}^g \textrm{Log} \left| c_{L_k} \right|^2 \cr
&&   + \frac{i}{2} \sum_{k=2}^g
\oint_{C_k} \left(  \textrm{Log}(\rho) - \frac{1}{2}\textrm{Log} |L'_k|^2   \right) \left( \frac{L''_k}{L'_k}dw - \frac{\bar{L''_k}}{\bar{L'_k}}d\bar{w}    \right)
\eea
where $F_d$ is some fundamental domain and $\rho(w)$ is the conformal factor that induces the hyperbolic structure on the metric $ds^2 = \rho(w,\bar{w}) |dw|^2 $. For genus greater than one, we need to find $\rho(w,\bar{w})$ such that the resulting geometry has Ricci scalar $R=-1$ which, upon writing $\rho(w,\bar{w})=e^{2\phi(w,\bar{w})}$, translates into the Liouville equation for $\phi$. For genus one, the metric is flat and so $\phi$ simply satisfies the $2D$ Laplace equation and the $\rho$ term is absent in \eqref{Liouvilleaction}. In fact for this case, since 
$$
L(w) = q w, L''(w) = c_L = 0,
$$
all the terms except for the first one in \eqref{Liouvilleaction} vanish, and we can compute $S_L=4\pi^2 \tau_2$ as stated earlier in equation \eqref{Ltorus}. More generally, finding the Liouville field amounts to finding a covering map $J$ that takes one from the upper half-plane $\mathds{H}$ to $\Omega$ since the Riemann surface $X$ can be uniformized by the action of some Fuchsian group on $\mathds{H}$. Thus, we have the following commutative diagram\footnote{See also \cite{Krasnov} for a nice exposition.}
\begin{center}
\begin{tikzpicture}[description/.style={fill=white,inner sep=2pt}] 
\matrix (m) [matrix of math nodes, row sep=3em, column sep=2.5em, text height=1.5ex, text depth=0.25ex] { \mathds{H} & & \Omega \\
& X & \\ };
\path[->,font=\scriptsize] 
(m-1-1) edge node [above] {$ J $} (m-1-3)
edge node [below left] {$ \pi_f $} (m-2-2) 
(m-1-3) edge node [below right] {$ \pi_s $} (m-2-2);
\end{tikzpicture}
\end{center}
where $\pi_f : \mathds{H} \rightarrow X$ and $\pi_s : \Omega \rightarrow X$ are the quotient maps corresponding to the Fuchsian and Schottky groups respectively. Taking the line element of $\mathds{H}$
to be $ds^2=|dz|^2/(\textrm{Im}(z)^2)$, the Liouville field $\phi(w,\bar{w})$ is then related to the map $J$ by
$
\phi(w,\bar{w}) = \textrm{Log} \left(\partial_w(J^{-1})(w)/ \textrm{Im} J^{-1}(w) \right).
$

\end{document}